        \documentstyle[]{l-aa-dipastro}
        \input psfig.sty

        \def\oneskip{\vskip 8pt}	
        \def\smallskip{\vskip 6pt}
        
\begin{document}

	\thesaurus{06(08.05.3; 08.16.4; 02.14.1; 08.01.1; 08.03.1; 09.16.1)}

        \title{TP-AGB stars with envelope burning}

	\author{ P. Marigo$^{1}$, A. Bressan$^{2}$, \and C. Chiosi$^{3,1}$}
	\institute{
                  $^1$ Department of Astronomy, University of Padova,
      Vicolo dell'Osservatorio 5, 35122 Padova, Italy \\        
                  $^2$ Astronomical Observatory, Vicolo dell'Osservatorio 5, 
                   35122 Padova, Italy \\
                  $^3$ European Southern Observatory, K-Schwarzschild-strasse 2,
      D-85748, Garching bei M\"unchen, Germany\\ }

	\offprints{P. Marigo  }
	\date{Received 2 July 1997; accepted 20 October 1997}

	\maketitle
	\markboth{Envelope Burning }{}

\begin{abstract} 

In this paper we focus on the TP-AGB evolution of intermediate-mass
stars  experiencing envelope burning ($M = 4\div5M_{\odot}$).
 Our  model of the TP-AGB phase 
is suitably designed to follow the 
peculiar behaviour of these stars, to which the simple analytical
treatment valid in the low-mass range can no longer be applied.

The approach we have adopted is a semi-analytical one as it combines
analytical relationships derived from complete models of TP-AGB stars
 with sole envelope models in which the physical structure is calculated
from the photosphere down to the core.
The solution for the envelope models 
stands on an original numerical method which allows to treat 
 major aspects of envelope burning. 

The method secures that,
during the quiescent inter-pulse periods, fundamental quantities such as
 the effective temperature, the surface luminosity, the
physical structure of the deepest and hottest layers of the envelope,
and the related energy generation from nuclear burning, are
not input parameters but the consequence of envelope model calculations.
This minimizes the use of analytical
relations,
thus giving our results greater homogeneity
and accuracy.

Moreover, we would like to draw 
the attention on the 
general validity of our algorithm which  
 can be applied also to 
the case of low-mass stars, in which envelope burning does not occur.

Our efforts are directed to analyse
the effects produced by envelope burning, such as: i) the
energy contribution which may drive significant deviations from the
standard core mass-luminosity relationship; and ii)  the changes in the
surface chemical composition due to nuclear burning via the CNO cycle.

Evolutionary models for stars with initial mass of $4.0, 4.5, 5.0 M_{\odot}$
and two choices of the initial chemical composition ($[Y=0.28, Z=0.02]$ and
$[Y=0.25, Z=0.008]$) are calculated from the first thermal pulse till the
complete ejection of the envelope. We find that massive TP-AGB stars can
rapidly reach high luminosities ($-6 > M_{\rm bol} > -7$), without exceeding,
however,  the classical limit to the AGB luminosity of $M_{\rm bol}\simeq
-7.1$ corresponding to the Chandrasekhar value of the core mass.
No carbon stars brighter than $M_{\rm bol} \sim -6.5$ are predicted to form
(the alternative of a possible transition from M-star to C-star during the
final pulses is also explored), in agreement with observations which indicate
that most of the very luminous AGB stars are oxygen-rich.

Finally, new chemical yields from  stars in the mass range 
$4 \div 5 M_{\odot}$ are 
presented, so as to extend the sets of stellar yields from low-mass
stars already calculated by Marigo et al. (1996). For each CNO element 
we give both
the secondary and the primary components.


\end{abstract}

\section {Introduction}
This study deals with the theoretical evolution of TP-AGB stars 
undergoing envelope burning -- i.e. those in which the base of the 
convective envelope deepens
into high temperature regions so that H-burning via the CNO cycle can
occur. This circumstance is usually encountered in AGB stars with
massive and extended envelopes ($M_{\rm env} \ga 2.5M_{\odot}$), as
indicated by several authors (Sugimoto 1971; Uus 1972; Iben 1973;
Renzini \& Voli 1981; Scalo et al. 1975; 
Bl\"ocker \& Sch\"onberner 
1991; Lattanzio 1992; Boothroyd \& Sackmann 1992; Vassiliadis \& Wood
1993; Bl\"ocker 1995a; Marigo et al. 1996). The most notable signature of
the TP-AGB evolution of these stars is
the break-down of the core mass-luminosity relation at high luminosities
(Bl\"ocker \& Sch\"onberner 
1991; Lattanzio 1992; Boothroyd \& Sackmann 1992).

Previous analyses of the AGB phase performed with the aid of 
envelope models 
have already pointed out that nuclear burning can take place in
the deepest and hottest layers 
of the convective envelope
(Scalo et al. 1975; Renzini \& Voli 1981; Marigo et al. 1996).

In Marigo et al. (1996) we adopted the 
assumption
that those regions do not provide  energy to the star, and
the constraint
that the core mass-luminosity relation is always satisfied.

The notable improvement of the present approach is to
account for the energy generated by the burning regions of the envelope.
For this purpose, we develop an
algorithm to include nuclear burning in the envelope of  a star 
in thermal equilibrium.

Establishing how deep in the interior
the convective envelope penetrates is crucial to determine  
the efficiency of the nucleosynthesis processes, for the
extreme sensitivity of nuclear rates to temperature and density. 
Furthermore, 
nuclear burning changes not only the chemical composition of the envelope,
but also affects the luminosity, the effective temperature, and the
 radius of the star. Therefore, the calculation of the envelope models must 
be carried out with great accuracy. 

The theoretical approach developed in this study offers
some advantages: 
it keeps the agility of purely
analytical models for quick 
computing and easy testing of different prescriptions but
 it has greater accuracy and 
self-consistency, which draw it closer to full calculations
of the TP-AGB phase.

This paper is organized in three main sections.
In the first part (Sect.~\ref{lmc}) we summarize 
the physical conditions under which the core-mass luminosity relation
is found and we analyze the 
break-down of this   due to envelope burning 
contributing  to the total stellar luminosity 
over the quiescent inter-pulse period (Sect.~\ref{overl}).
The theoretical method developed to this aim is described and 
discussed (Sect.~\ref{numer}).

In the second part (Sect.~\ref{evol}) we apply our method to follow 
the TP-AGB phase of the $4.0, 4.5, 5.0 M_{\odot}$ stars with two choices 
of the initial chemical composition, namely  $[Y=0.25, Z=0.008]$ and 
$[Y=0.28, Z=0.02]$. The calculations
are carried out from the end of the E-AGB up to the complete ejection of
the envelope. 

In the third part (Sect.~\ref{nucl}) we study the nucleosynthesis associated to
envelope burning which, together with the third dredge-up, concurs 
to alter the surface abundances of the elemental species.
Finally (Sect.~\ref{yields}),
we provide new chemical yields from  stars in the above mass range, 
  distinguishing for each element of the  CNO group
the primary and secondary contribution.

\section{The core mass-luminosity relation}
\label{lmc}
Paczynski (1970) first discovered the existence of a simple, almost linear
 relationship between the core mass of
a double-shell burning star and its quiescent inter-flash luminosity along the
TP-AGB. 

Over the years, continuous upgrading of the input physics
(e.g. new opacities, revised nuclear reaction rates, 
equations of state, etc.) and 
detailed calculations of the  AGB phase for wide ranges of stellar masses
and chemical compositions resulted into a flourishing of different core 
mass-luminosity ($M_{\rm c} - L$)
relationships (Iben 1977; Iben \& Truran 1978; Havazelet \& Barkat 1979;
Wood \& Zarro 1981; Lattanzio 1986; Boothroyd \& Sackmann 1988a).

In this work (see also Marigo et al. 1996) we adopt the following
relations: 

\begin{equation}
\label{lmc1}
L_{M_{\rm c}} = 238\,000 \mu^{3} Z^{0.04}_{\rm CNO} (M_{\rm c}^2 - 0.0305 M_{\rm c} -
0.1802) 
\end{equation}

\noindent
for stars with  core mass in the range 
$0.5 M_{\odot} \le M_{\rm c} \le 0.66 M_{\odot}$ 
(Boothroyd \& Sackmann 1988a);
and 

\begin{equation}
\label{lmc2}
L_{M_{\rm c}} = 122585 \mu^2 (M_{\rm c} - 0.46) M^{0.19}
\end{equation}

\noindent
for stars with core mass $M_{\rm c} \ge 0.95 M_{\odot}$ 
(originally
from Iben \& Truran 1978 according to the slightly modified version of 
Groenewegen \& de Jong 1993;
see also Iben 1977).

In these formulas stellar masses $M$ and luminosities $L$ are in solar units;
$Z_{\rm CNO}$ is the total abundance (in mass fraction) of C, N, O isotopes in
the envelope; $\mu = 4/(5X + 3 - Z)$ is the mean molecular weight under the
assumption of a fully ionized gas, where $X$ and $Z$ are the abundances of
hydrogen and metals, respectively. For stars with core mass  
$0.66 M_{\odot} < M_{\rm
c} < 0.95 M_{\odot}$ a linear interpolation is performed.

In our model we also account for the steep luminosity increment as a
function of the core mass during the first inter-pulse periods, before the
onset of the ``full amplitude'' regime to which the standard $M_{\rm c}-L$
relation given by Eqs.~(\ref{lmc1}) or (\ref{lmc2}) refers. 
The initial subluminous evolution  
is followed by adopting a simple linear relation
between the luminosity and the core mass,
$L_{\rm first} = S \times M_{\rm c} + Q$, with the slope $S$ given
as a function of the initial mass $M_{\rm i}$ of the star:

\begin{equation}
S = 60761.2 \times {\rm exp}(M_{\rm i}/2.)
\label{lfirst}
\end {equation}  

\noindent
where masses and luminosities are expressed in solar units.
This relation is a least-square fit to the results presented 
by Vassiliadis \& Wood (1993) (see their Fig.~12).
For each value of $M_{\rm i}$, the intercept $Q$ is 
fixed by the values of the luminosity and core mass at the first thermal pulse.

The existence of the core mass-luminosity relation for AGB stars was given a
transparent analytical explanation by Tuchman et al. (1983) just
starting from the equations of stellar structure under specific physical
conditions (``radiative zero solution''; see Eggleton 1967; Paczynski 1970).
The stellar structure is required to possess a degenerate core 
of mass $M_{\rm c}$ surrounded by a
narrow radiative shell (or double shell) source providing the luminosity,
beyond which there  must exist a thin 
(with a mass $\Delta M \ll M_{\rm c}$) and inert 
(the luminosity is constant) {\it transition region} in
radiative equilibrium. This is bounded at the top by the base of the convective
envelope.
Integrating the equations of hydrostatic and radiative equilibrium over the
{\it transition region}, after simple substitutions, Tuchman et al. (1983)
derived an analytical expression relating the luminosity $L$ of the star to
its core mass $M_{\rm c}$, with some dependence on the chemical composition
and the temperature just above the burning shell. The effect of the outer
envelope on this relation is 
negligible because of the extremely steepness of pressure,
temperature, and density gradients in the radiative zone.
Thus, during the quiescent TP-AGB phase, the evolution of the core is expected
to be essentially decoupled from that of the envelope (no mass dependence).

However, Tuchman et al. (1983) suggested that the $M_{\rm c}
- L$ relation is likely to break down for stars with envelope burning,
as the base of the 
convective envelope may penetrate so deeply as to eat up the radiative
buffer over the H-burning shell.

Such a possibility was anticipated by Scalo et al. (1975) who pointed out 
that, during the quiescent inter-pulse period,
in massive and luminous AGB stars the
envelope convection may extend all the way down to the H-burning shell, as a 
result of the high radiation pressure $P_{\rm R}$.
In brief, when the contribution of $P_{\rm R}$ to the total pressure
becomes significant, the radiative ($\nabla_{\rm rad}$) and adiabatic 
($\nabla_{\rm ad}$) temperature gradients are expected 
to decrease less steeply with depth, and to reach the minimum value of 
0.25, respectively.
Both facts concur in drawing the base of the convective envelope 
-- reached when $\nabla_{\rm rad}= \nabla_{\rm ad}$ -- deeper in the interior.
High temperatures ($T_{\rm b} > 40 \times 10^{6}$ K) can be attained,
thus allowing nuclear burning at the base of the envelope.

\section{Envelope burning as an energy source}
\label{overl}
Over the years it has become clear that the energy generation due to
envelope burning can make massive TP-AGB stars to significantly deviate from
the $M_{\rm c} - L$ relation. High luminosities are soon reached
($M_{\rm bol} < -6$), at
least until the residual envelope is massive enough
to allow for high temperatures ($T_{\rm b} > 40 \times 10^{6}$ K) 
at its  bottom.  All this is indicated by complete 
models of  TP-AGB stars (Bl\"ocker \&
Sch\"onberner 1991; Lattanzio 1992; Boothroyd \& Sackmann 1992; Vassiliadis \&
Wood 1993). 

Earlier analyses, primarily aimed to study the structure of envelope burning
 and associated nucleosynthesis, were carried out with the aid of 
convective envelopes models (Scalo et al. 1975;
Renzini \& Voli 1981). The same kind of theoretical approach adopted in the 
semi-analytical TP-AGB model by Marigo et al. (1996). The assumption common to 
all those models is that the envelope satisfies the classical core
mass-luminosity relation. This is no longer valid  in massive
AGB stars, due to the energy generation by envelope burning.
Moreover, numerical integrations were made under the constraint that the
radial coordinate of the base of the convective envelope coincides with
the core radius (Renzini \& Voli 1981; Marigo et al. 1996). This
approximation can be risky in the case of  massive TP-AGB stars
as the base of the convective envelope falls in a region, which is 
very  thin in mass but across which the structural gradients are 
very steep. Therefore, fixing {\it a priori} the inner extension of
the envelope convection may lead to heavily under- or over-estimate  
the base temperature, to the value of which the nuclear rates are highly
sensitive. 
In contrast,  the correct treatment of the envelope burning requires that
the energy being generated by nuclear reactions is taken into account
and the exact location of the deepest layer through which external convection 
penetrates is determined (e.g. by the Schwarzschild condition).

\subsection{The method of solution}
\label{numer}

First of all, we relax the original assumption 
fixing the local luminosity at a constant value (equal to $L$)
throughout the envelope (i.e. no energy source).
This condition is replaced with the equation of energy balance:

\begin{equation}
\label{energ}
\frac{dL_{\rm r}}{dr} = 4 \pi r^2 \epsilon_{\rm r} \rho_{\rm r}
\end{equation}

\noindent
where $L_{\rm r}$ is the local luminosity;
$\rho_{\rm r}$ the gas density; $\epsilon_{\rm r}$ is the amount
of energy released
by nuclear transmutation of hydrogen
into helium per unit of mass and time. All the variables are evaluated
at the radius $r$.

The integration of Eq.~(\ref{energ}) across the envelope
must satisfy  the basic constraint  given by the principle of
energy conservation. This means that during
the quiescent inter-pulse period the surface luminosity $L$
is the sum of different contributions, namely those from:
the radiative H-burning shell, $L_{\rm H}$;
the nuclear burning at the base of the convective envelope, $L_{\rm EB}$;
the small contributions from the He-burning shell, $L_{\rm He}$, 
the gravitational contraction of the core, $L_{\rm G}$, and the rate of energy
loss via neutrinos, $L_{\rm \nu}$:

\begin{equation}
\label{lumeq}
L  = \int_{0}^{R_{\rm S}} \frac{dL_{\rm r}}{dr} dr 
	   =  L_{\rm G}+L_{\rm He} + L_{\rm H} + L_{\rm EB} - L_{\rm \nu}
\end{equation}

\noindent
where $R_{\rm S} = \sqrt{L/(4 \pi \sigma T_{\rm eff}^{4})}$
is the stellar surface radius.
In this study we do not consider neutrino losses.
This negative contribution is very small compared to the total
luminosity, and it is indeed negligible in low-mass stars (Iben \& Tutokov
1984; K\"oster \& Sch\"onberner 1986).
In
massive cores (i.e. $M_{\rm c} \ga 0.8 M_{\odot}$), its absolute value may
be initially comparable to the gravitational energy release
from core contraction during the quiescent inter-pulse period  (Bl\"ocker
1995a,b). However, as the star evolves the core
interiors become cooler and more degenerate so that 
the neutrino luminosity falls again well below 
the gravothermal luminosity.

Each term  (but for $L_{\rm \nu}$) of the sum in Eq.~(\ref{lumeq}) can be
explicitly expressed as:

\begin{eqnarray}
\label{lterms}
L_{\rm G}+L_{\rm He} & = & \int_{0}^{R_{\rm core}} \frac{dL_{\rm r}}{dr} dr \nonumber \\
L_{\rm H} & = & \int_{R_{\rm core}}^{R_{\rm conv}} \frac{dL_{\rm r}}{dr} dr \\
L_{\rm EB} & = & \int_{R_{\rm conv}}^{R_{\rm S}} \frac{dL_{\rm r}}{dr} dr \nonumber
\end{eqnarray}

\noindent
where $R_{\rm core}$ is the core radius,
here defined as the radius at the bottom boundary of the H-burning
shell, and
$R_{\rm conv}$ is the radial coordinate of the base of the 
convective envelope.
In the system of equations~(\ref{lterms})
the only quantity that can be  directly evaluated
with our envelope model is $L_{\rm EB}$, whereas for 
the other terms  we need some analytical prescriptions.
We make use, in particular, of the relation

\begin{equation}
L_{\rm G} + L_{\rm He} =  2000 (M/7)^{0.4} {\rm exp}[3.45 (M_{\rm c} - 0.96)]
\end {equation}

\noindent
derived from the results of complete
calculations of the AGB phase 
(Iben 1977; see also Groenewegen \& de Jong 1993). 

\begin{figure}
\psfig{file=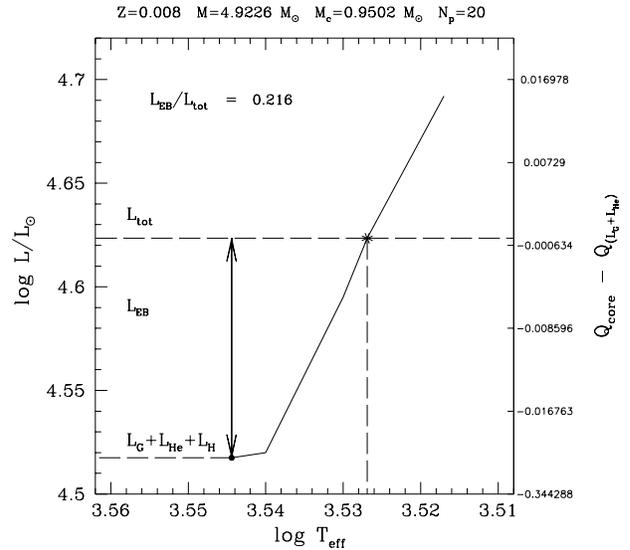,width=8.5truecm}
\caption{ Solution of the system of two equations~(\protect\ref{sys}) in 
the plane $L-T_{\rm eff}$. The solid line corresponds to the locus
where $F(L,T_{\rm eff})=0$ ($L(R_{\rm conv})=L_{\rm M_{\rm c}}$);
the asterisk marks the point (solution) where 
$G(L,T_{\rm eff})=0$ ($Q_{\rm core} = Q_{ (L_{\rm He}+L_{\rm G}) }$).  }
\label{meth}
\end{figure}

As a first approximation, we consider the energy contribution from envelope
burning, $L_{\rm EB}$, as a correction term which is
added to the luminosity $L_{\rm M_{\rm c}}$ 
predicted by the $M_{\rm c}-L$ relation,   Eqs.~(\ref{lmc1}) and (\ref{lmc2}),
to get the 
total luminosity, i.e.
$L \sim L_{\rm M_{\rm c}} + L_{\rm EB}$.
Substituting $L$ into Eq.~(\ref{lumeq}), we can express
the luminosity provided by the radiative H-burning shell as:

\begin{equation}
\label{lh}
L_{\rm H} = L_{\rm M_{\rm c}} - L_{\rm G} - L_{\rm He} 
\end{equation}

There is one aspect of the above scheme deserving some remarks. We have 
assumed that $L_{\rm H}$ and $L_{\rm EB}$ are decoupled, whereas in 
a real star with envelope burning as soon as the envelope convection
penetrates into the H-burning region the two luminosities 
are related to each other in a way that cannot be ascertained {\it a priori}
without calculating a complete stellar model.
A possible way out of this difficulty could be a scheme in which 
the structure of the
radiative double-shell source  with ``deep envelope'' is integrated
down to
the carbon-oxygen core. This approach could be explored in a future study.
On the other hand, a preliminary analysis shows that the relative error of
our approximation for $L_{\rm H}$ (see Eq.~(\ref{lh})) with respect to the
results from full AGB calculations is mostly comprised within $\sim 2 \% \div
10 \%$ (Marigo 1998, in preparation).

However, despite the adopted
approximation, the present model is a step forward when compared
with previous analyses of envelope burning performed with the aid
of purely analytical assumptions (Groenewegen \& de Jong 1993), or
static envelope models (Marigo et al. 1996).
Moreover, the results obtained with this scheme turn out to be consistent 
with those from full evolutionary calculations.

\subsection{Boundary conditions}

\begin{figure}[t]
\psfig{file=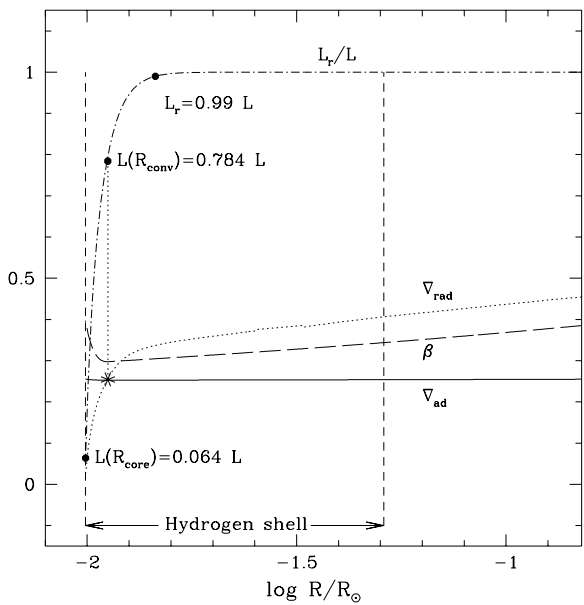,width=8.5truecm}
\caption{Physical conditions in the deepest layers of the model
corresponding to the case illustrated in Fig. \protect\ref{meth}.
The variables $\nabla_{\rm rad}$, $\nabla_{\rm ad}$, $\beta=P_{\rm gas}/P$,
$L_{r}/L$ are plotted as function of the radial coordinate.
The arrows indicate the width of the H-burning shell, with 
the upper boundary
defined by an energy generation rate of 10 erg gr$^{-1}$ s$^{-1}$.
A few values of the fractional luminosity $L_{\rm r}/L$ are displayed at
interesting points (marked by full circles), namely:
i) the inner boundary of the H-burning shell at $R_{\rm core}$;
ii) the base of the convective envelope 
(when $\nabla_{\rm rad}=\nabla_{\rm ad}$,
asterisk)
at $R_{\rm conv}$; and iii) the level at which the local luminosity
amounts to the $99\%$ of the total luminosity.
In this case the envelope convection extends deeply in the
interior of the H-burning shell.}
\label{hshell1}
\end{figure}

\begin{figure}[t]
\psfig{file=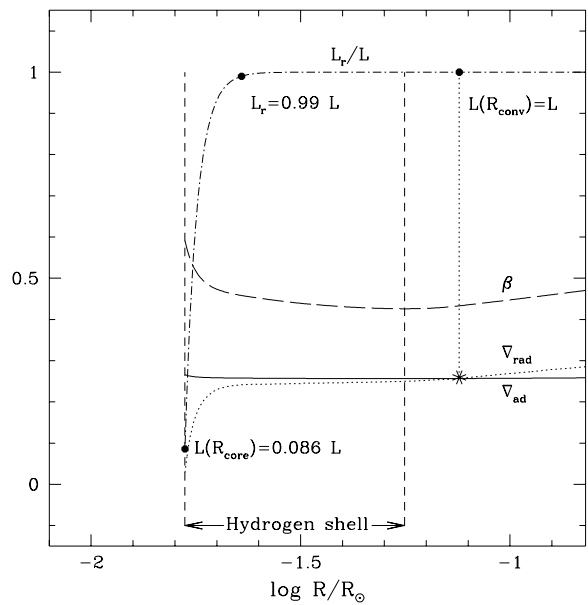,width=8.5truecm}
\caption{The same as in Fig. \protect\ref{hshell1}, but
referred to the stage soon after the end of the E-AGB phase.
In this case a radiative {\it transition region} exists
between the upper boundary of the H-burning shell and the base of the 
convective envelope so that the core mass-luminosity relation still holds.}
\label{hshell2}
\end{figure}

Given the stellar mass $M$, the core mass $M_{\rm c}$, the chemical
composition of the envelope, the luminosities $L_{\rm H}$, $L_{\rm
He}$, $L_{\rm G}$ at a certain time, the task is to find 
 $T_{\rm eff}$ and $L_{\rm EB}$ (and hence total
luminosity $L$) of the stellar model satisfying 
Eq.~(\ref{lumeq}). The solution is univocally singled out by means of envelope
integrations constrained by two boundary conditions, which naturally
derive from the system of equations~(\ref{lterms}) :

\begin{eqnarray}
\label{vcore}
L (R_{\rm core}) & = & L_{\rm G} +L_{\rm He}  \\
\label{vconv}
L (R_{\rm conv}) & = &  L_{\rm G} +L_{\rm He} + L_{\rm H} 
\end{eqnarray}

We define the variable $Q(r)=M(r)/M$ as the mass coordinate of the layer
of radius $r$, where $M(r)$ is the mass contained below it
and $M$ is the total stellar mass.
Let us call $Q_{\rm core}$ and $Q_{(L_{\rm He}+L_{\rm G})}$ the
mass coordinates of the core and of the layer with a luminosity
$L_{\rm He}+L_{\rm G}$, respectively.
According to the adopted definition of the core,
it is physically equivalent to replace Eq.~(\ref{vcore}) with:

\begin{equation}
\label{qcore}
Q_{\rm core} = Q_{(L_{\rm He}+L_{\rm G})} 
\end{equation}

\noindent
which is more suited to our calculations.
To summarize, we deal with two boundary conditions 
(Eqs.~(\ref{vconv}) and (\ref{qcore})),
which require to solve a system of two equations in the two unknowns
$L$ and $T_{\rm eff}$:

\begin{equation}
\left\{ \begin{array}{lllll}
G(L,T_{\rm eff}) & = & Q_{\rm core} - Q_{(L_{\rm He}+L_{\rm G})} = 0 \\
F(L,T_{\rm eff}) & = & L (R_{\rm conv}) - (L_{\rm G} + L_{\rm He} + L_{\rm H}) = 0 
\end{array}
\label{sys}
\right.
\end{equation}

The adopted numerical technique is Brent's method for root finding (Numerical
Recipes 1990) applied to both functions.
A tolerance $\epsilon_{ L}=10^{-5}$ on the root of function $F$ 
(expressed as the difference of logarithmic luminosities),
seems reasonable for our purposes.
Because of the extreme steepness of the structural variables across the
H-burning shell, we need to work with a high accuracy, 
$\epsilon_{Q} = 10^{-6}$, on the root of function $G$, 
such that iterations will continue until
$\left| Q_{\rm core} - Q_{(L_{\rm He}+L_{\rm G})}\right | < \epsilon_{Q}$.

This  method proves to be  general and able to work also when
envelope burning is not operative, e.g. 
in the final stages of high mass stars
when the mass of the residual envelope is
drastically reduced, or in the whole TP-AGB 
evolution of low-mass stars which never deviate from the $M_{\rm c}-L$
relation. 
In these cases $L_{\rm EB}=0$ so that 
the local luminosity all over the envelope coincides 
with the surface luminosity given by $L=L_{\rm M_{\rm c}}$.
The only unknown quantity is then $T_{\rm eff}$ which is the root
of the function $G(L_{\rm M_{\rm c}}, T_{\rm eff})$. 

In order to show how the method works, we consider the case
of  a stellar model with current mass $M = 4.9226 M_{\odot}$, 
core mass $M_{\rm c} = 0.9502 M_{\odot}$, and luminosity
$\log L_{\rm M_{\rm c}} = 4.5175$, this latter being derived from
the $M_{\rm c} - L$ relation.
These values refer the final stage of the $20^{\rm th}$ inter-pulse
period  of  a 
$5 M_{\odot}$, $[Y=0.25, Z=0.008]$ star.
In the $\log L - \log T_{\rm eff}$  plane of  Fig. \ref{meth} we 
show the locus (solid line) of the models satisfying the condition  
$F=0$ of Eq.~(\ref{vconv}). Each $i^{\rm th}$ point along the line
is derived from envelope integrations carried out 
with an input luminosity 
$L_{i} = (L_{\rm G}+L_{\rm He}+L_{\rm H}) + \Delta L_{i}$,
where $\Delta L_{i}$ is an arbitrary amount of extra-luminosity.
At given $L_{i}$ there is a unique value of the effective temperature, 
$T_{{\rm eff},i}$
(hence a unique envelope model) root of the function $F$.
Along the right-hand side vertical axis, several values of the function
$G$ -- e.g. $G(L_{i}, T_{{\rm eff},i})$ -- derived from
envelope models satisfying the condition $F=0$ are displayed in a
one-to-one correspondence with the luminosity. 

The solution of the system of
 Eqs.~(\ref{sys}) is represented by the pair of values ($\log L= 4.62342 $,
$\log T_{\rm eff}= 3.52689$) at which both $F=0$ and $G=0$. This is 
indicated by the asterisk along the solid line.
In this particular model,  the
contribution from envelope burning, $L_{\rm EB}$, amounts to $\sim
22\%$ of the total luminosity.

Figure~\ref{hshell1} shows a few physical quantities characterizing 
the structure of this model in the deepest regions of the
envelope and beyond, down to the lower
boundary of the H-burning shell, $R_{\rm core}$. 
To be noticed is that the base of the convective
envelope is reached in the deep interior of the H-burning shell,
where the temperature is $T_{\rm b} \sim 100 \times 10^{6}$.

For comparison, the same quantities are plotted in Fig. \ref{hshell2}, but 
relative to the $5 M_{\odot}$, $[Y=0.25, Z=0.008]$ star soon after the end of
the E-AGB phase, with $M_{\rm c}=0.9344 M_{\odot}$, $\log T_{\rm eff}=3.5583$,
and $L_{\rm M_{\rm c}} = 4.3502 L_{\odot}$.
At this stage, envelope burning is very weak ($T_{\rm b} \sim
24 \times 10^{6}$; $L_{\rm EB}=0$), so that 
the star still obeys the $M_{\rm c} - L$ relation.
Note the existence of the radiative buffer between the H-burning shell and
the base of the convective envelope.

\section{Evolutionary calculations}
\label{evol}

With the method described in Sect. \ref{numer}
we calculated the TP-AGB evolution of stars with initial mass of  
$4.0, 4.5, 5.0 M_{\odot}$ and two choices of the chemical composition 
$[Y=0.28, Z=0.02]$ and $[Y=0.25, Z=0.008]$.
In the following we briefly recall the basic ingredients adopted in the
model, the same ones employed by Marigo et al. (1996) to whom the redear should
refer for more details.

The initial conditions at the first thermal pulse are taken from full
evolutionary calculations extending from the main sequence till the beginning
of the TP-AGB phase kindly provided us by L\'eo Girardi (1997, private
communication). For each stellar mass we single out the first significant pulse,
usually defined as the first He-shell flash in which $L_{\rm He}^{\rm max} >
L$, where $L$ is the surface luminosity (Boothroyd \& Sackmann 1988b).
All the quantities of interest are then referred to the
time corresponding to the quiescent pre-flash
luminosity maximum in the light curve.
This procedure let us  account for the changes in the surface
chemical composition caused by the  previous mixing processes, namely,
the first dredge-up as the star first reaches its Hayashi track
after central hydrogen exhaustion, and the second
dredge-up at the base of the E-AGB
(Tables~\ref{dred12z008} and \ref{dred12z02}).

It is worth noticing that the input physics of our envelope model is the same 
as in the Padua  stellar evolution code, so that our results on the TP-AGB
phase are homogeneous with
the previous evolutionary history. The main characteristics of these stellar
models can be summarized as follows:

\begin{itemize}
\item{
The stellar models in usage here 
to derive the initial conditions allow for 
convective overshoot from the core and external envelope 
 (Alongi et al. 1993; see also Marigo et al. 1996). 
Therefore, 
the maximum initial mass, $M_{\rm up}$, for a star to build an 
electron-degenerate C-O core, i.e. to evolve through the AGB phase,
is around $5 M_{\odot}$.  Classical models without overshoot
predict a higher value, i.e. $M_{\rm up} \sim 8 M_{\odot}$.}

\item{
The high-temperature opacities are those
of Iglesias \& Rogers (1996). They are particularly suitable to AGB
calculations inclusive of the third dredge-up as the authors also provide
tables for a variety of carbon- and oxygen-rich mixtures.
At low temperatures ($T < 10^{4} K$)
we use the molecular opacities of Alexander \& Ferguson (1994). }

\item{
In the outermost super-adiabatic convective region,
the adopted value of the mixing length parameter is $\alpha =
1.68$, according to the calibration of the solar model. }

\item{ Finally, the CNO-cycle nuclear reaction rates are 
from Caughlan \& Fowler (1988).}
\end{itemize}

The analytical relations used to construct our TP-AGB synthetic model are:

(1) {\it The core mass-luminosity relation}, given by Eq.~(\ref{lmc1}) or Eq.~
(\ref{lmc2}).

(2) {\it The core mass-interpulse period relation}, taken from 
Boothroyd \& Sackmann (1988b):
\begin{equation}
\log T_{\rm ip}  = \left\{
\begin{array}{ll}
      4.50 \: (1.689 - M_{\rm c}) \:\:\:\:\:\:\:\: {\rm for} \:\:\: Z=0.02 \\
      4.95 \: (1.644 - M_{\rm c}) \:\:\:\:\:\:\:\: {\rm for} \:\:\: Z=0.001 
\label{mctip}
\end{array}
\right.\ 
\end{equation}
where $T_{\rm ip}$ and $M_{\rm c}$ are expressed in years and solar units,
 respectively.

(3) {\it The rate of evolution}, according to Groenewegen \& de Jong (1993):
\begin{equation}
\frac{d M_{\rm c}}{dt} = 9.555 \: 10^{-12} \: \frac{L_{\rm H}}{X} 
\label{evrate}
\end{equation}
where $d M_{\rm c}/dt$ is the growth rate of the core mass 
($M_{\odot} yr^{-1}$), $X$ is the hydrogen
abundance (in mass fraction) in the envelope, $L_{\rm H}$ is the luminosity
produced by H-burning (in solar units). The numerical factor takes into account
the energy
released from the nuclear conversion of $1\: g$ of hydrogen into 
helium ($\sim 6.4\:\:10^{18}\: erg $).

(4) {\it Flash-driven luminosity variations} --
We apply proper corrections as a function of the envelope mass
(Groenewegen \& de Jong 1993) to 
account for the deviations in the luminosity caused by the occurrence 
of thermal pulses, i.e. the post-flash luminosity peak and 
the low-luminosity dip.

(5) {\it Mass loss} --
The adopted prescription for the mass-loss rate 
is the semi-empirical one of Vassiliadis \&
Wood (1993), which gives the rate as a function of the pulsational period of
variable AGB stars:
\begin{equation}
\log \dot M = -11.4 + 0.0123 P
\label{mlr1}
\end{equation}
\begin{equation}
\dot M =  6.07023 \:\:10^{-3}\frac{L}{c v_{\rm exp}}, 
\label{mlr2}
\end{equation}
Here, $\dot M$ is given in units of $M_{\odot}$ yr$^{-1}$, the stellar
luminosity $L$ is expressed in $L_{\odot}$, the pulsation period $P$ in days,
$c$ is the light speed (in km s$^{-1}$) and $v_{\rm exp}$ (in km s$^{-1}$)
denotes the terminal velocity of stellar wind. During calculations, at any time
step, the mass-loss rate is chosen to be the minimum value between those
given by Eq. (\ref{mlr1}) and (\ref{mlr2}). The wind expansion velocity
$v_{\rm exp}$ (in km s$^{-1}$) is estimated as a function of the period of
pulsation:
\begin{equation}
v_{\rm exp} = -13.5 + 0.056 P
\end{equation}
with the additional constraint that $v_{\rm exp}$ lies in the range
$3.0 - 15.0 $ km s$^{-1}$, the upper limit being the typical terminal velocity
detected in high mass-loss rate OH/IR stars.

The pulsation period $P$ is derived from the period-mass-radius relation (Eq.
(4) in Vassiliadis \& Wood 1993), with the assumption that variable AGB stars
are pulsating in the fundamental mode:
\begin{equation}
\log P = -2.07 + 1.94\:\: \log R - 0.9\:\: \log M
\label{periodo}
\end{equation}
where the period $P$ is given in days; the stellar radius $R$ and mass $M$ are
expressed in solar units.

For each stellar mass the calculations are carried out starting from the first
significant pulse until the  envelope mass, $M_{\rm env}$ reduces to about
$0.001 M_{\odot}$ that hydrodynamical studies suggest to be the critical
value at which the pulsational instability (and hence strong mass-loss) ceases
(Tuchman et al. 1979). It is worth recalling that full evolutionary
calculations show that a model star moves off the AGB when the residual
envelope mass is diminished below $0.01 M_{\odot}$ (Sch\"onberner 1983).
From the present results it turns out that the reddest point 
in the H-R diagram (see Fig.~\ref{hrd})
is reached when $M_{\rm env} \sim 0.8 \div 0.9 M_{\odot}$ in all cases studied,
 with a typical value $\log T_{\rm eff} \sim 3.48$ for models with
$Z=0.008$, and  $\log T_{\rm eff} \sim 3.44$ for models with
$Z=0.02$.
Then, the tracks slowly move toward the blue part of the H-R diagram,
lying in the interval $[3.44, 3.48 + 0.03]$ until $M_{\rm env}
 \sim 0.3 \div 0.4 M_{\odot}$.
The final evolution of the models is characterized by 
a quick decrease of $T_{\rm eff}$ (because of the very small envelope mass). 
However, we do not pretend here to handle the delicate topic about the
transition from the AGB to the formation of planetary nebulae, which is
beyond the aim of the present study.

The total number of thermal pulses included in each evolutionary
calculation is $N_{\rm P}^{\rm tot} = 174$ for the $[4.0 M_{\odot}, Z=0.008]$
model; $N_{\rm P}^{\rm tot} = 146$ for the $[4.5 M_{\odot}, Z=0.008]$ model;
$N_{\rm P}^{\rm tot} = 129$ for the $[5.0 M_{\odot}, Z=0.008]$ model.
A smaller number of thermal pulses are expected to be suffered by 
stars with the same initial mass, but lower (e.g. solar) metallicity.
 Our calculations yield
$N_{\rm P}^{\rm tot} = 39$ for the $[4.0 M_{\odot}, Z=0.02]$
model; $N_{\rm P}^{\rm tot} = 56$ for the $[4.5 M_{\odot}, Z=0.02]$ model;
$N_{\rm P}^{\rm tot} = 83$ for the $[5.0 M_{\odot}, Z=0.02]$ model.

\subsection{Evolution in the $M_{\rm c} - L$ plane}
\label{lumin}

\begin{figure}
\psfig{file=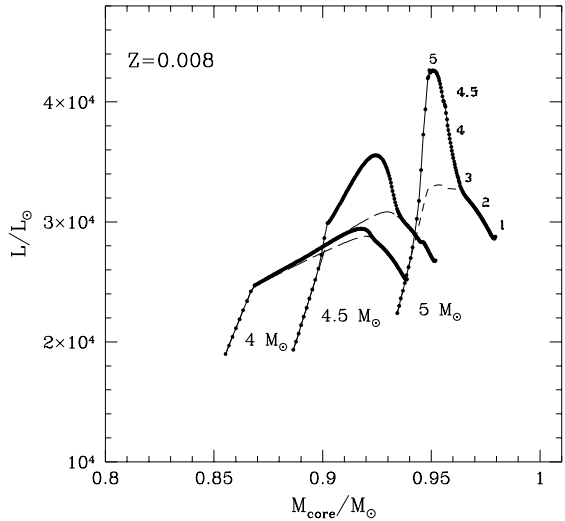,width=8.5truecm}
\caption{ 
Luminosity evolution of 4.0, 4.5, 5.0 $M_{\odot}$ TP-AGB stars with
composition  $[Y=0.25, Z=0.008]$ as
a function of the core mass. The dashed lines correspond to the standard core
mass-luminosity relation, the filled circles refer to the actual stellar
luminosity, at its maximum value before each pulse. The overluminosity
produced by envelope burning rapidly reaches a maximum value, it starts to
decrease after the onset of the super-wind (the numbers along the $5
M_{\odot}$ curve indicate the current stellar mass in solar units), finally it
vanishes and the star approaches again the core mass-luminosity relation.  }
\label{extralz008}
\end{figure}

\begin{figure}
\psfig{file=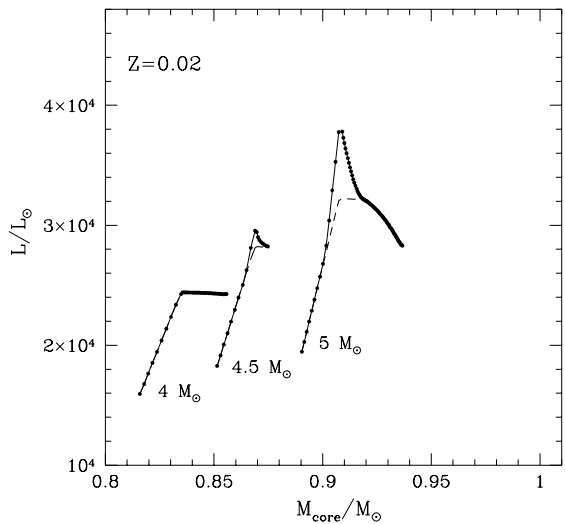,width=8.5truecm}
\caption{ 
The same as in Fig.~\protect\ref{extralz008}, but with $[Y=0.28, Z=0.02]$. }
\label{extralz02}
\end{figure}

As already mentioned, the energy contribution from envelope burning makes
massive TP-AGB stars to leave the $M_{\rm c}-L$ relation during part of their
evolution. 
This effect is illustrated in Figs.~\ref{extralz008} and \ref{extralz02}.
During the initial stages of the thermally pulsing regime the luminosity
evolution is governed by the H-burning shell, just re-ignited after the second
dredge-up. 
As the efficiency of the envelope burning
increases, the star can quickly reach higher and higher luminosities 
(steeply rising
parts of the  curves), with a rate much greater than expected from the
slope of the $M_{\rm c}-L$ relation (dashed lines).
The luminosity growth goes on as long as the envelope remains massive enough 
to allow high temperatures at the base of the envelope 
(see Sect.~\ref{tb} below).
When mass loss becomes significant (and envelope burning
weakens) the luminosity starts to decline and
the star eventually approaches again the $M_{\rm c}-L$ relation, where it lies
till the end of the AGB phase. This kind of luminosity evolution agrees
with the results from full evolutionary calculations of the TP-AGB phase
(Boothroyd \& Sackmann 1992; Vassiliadis \& Wood 1993).

The above trend may of course vary 
depending on several factors, part of which are intrinsic to the stars,
e.g. initial mass and chemical composition,  and
part to the theoretical model, e.g. mixing formalism,
mass-loss prescription, opacities, and nuclear reaction rates (see 
Renzini \& Voli 1981; Sackmann \& Boothroyd 1991; Boothroyd \& Sackmann
1992, 1993; Bl\"ocker 1995a).

Figures~\ref{extralz008} and \ref{extralz02} show the effect on the above trend
of mass and chemical composition. The deviation from the luminosity 
predicted by the $M_{\rm
c}-L$ relation is more pronounced in more massive stars at fixed metallicity.
Let us consider the $[Y=0.25, Z=0.008]$ 
case (Fig.~\ref{extralz008}). The maximum
relative contribution of envelope burning to the total luminosity 
$L_{\rm EB}/L $ goes from about $0.02$  for the $4.0 M_{\odot}$ model, 
$0.14$ for the $4.5 M_{\odot}$ model, up to  $0.23$ for the $5
M_{\odot}$ model. Concerning this latter, the plot shows how the
decline in luminosity  depends on the current value of the stellar mass
(indicated along the curve), as it is progressively reduced by stellar winds. 
At higher
metallicities (Fig.~\ref{extralz02}), the overluminosity generated by the
 envelope burning is much smaller at given initial mass. In the 
$[Y=0.28, Z=0.02]$ case,
the maximum relative contribution  is negligible  
for the $4.0 M_{\odot}$ model (which
actually never deviates from the $M_{\rm c}-L$ relation), and very small
for the $5.0$ and $4.5 M_{\odot}$ stars, where 
 $L_{\rm EB}/L$ amounts to $0.15$ and $0.05$, respectively.

Concluding this section, we like to briefly comment on the dependence 
of $M_{\rm c}-L$ relation given by Eq.~(\ref{lmc2}) on the current star mass 
(i.e. the factor $M^{0.19}$).
Iben (1977) originally introduced a factor $M^{0.4}$ to fit 
 the results for the AGB evolution (at constant mass) of his $7.0 
M_{\odot}$ star. Subsequently Iben \& Truran (1978) reduced the 
exponent to $0.19$ to obtain a compromise  with the results of 
Paczynski (1970).

This dependence is indeed quite weak, and it has in practice
no effect for most of the TP-AGB lifetime. However, 
after the onset of the super-wind, envelope stripping
gets so efficient that the weight of the mass factor in  Eq.~(\ref{lmc2})
becomes important. 
In Fig.~\ref{extralz008} the final, slow decrease in luminosity at
increasing core mass during the last evolutionary stages is just caused
by the drastic reduction of the stellar mass.

This trend seems to contradict the existence of
 the core mass-luminosity relation itself, because this latter 
implies that the luminosity should not depend on the envelope mass.
We may argue that the factor $M^{0.19}$  actually masked some 
low-efficiency  envelope burning (the adopted mixing length parameter
 was $\alpha=0.7$ in Iben's calculations, roughly corresponding to 
$\alpha \sim 1.4$ of other groups). Though a cautionary remark on the
use of such a relation is worthy  since it may depend on model details, 
the general applicability of the our solution scheme is not affected, and
the adoption of different $M_{\rm c}-L$ relations could be explored in a
future study.

\subsection{The base temperature}
\label{tb}
\begin{figure}[t]
\psfig{file=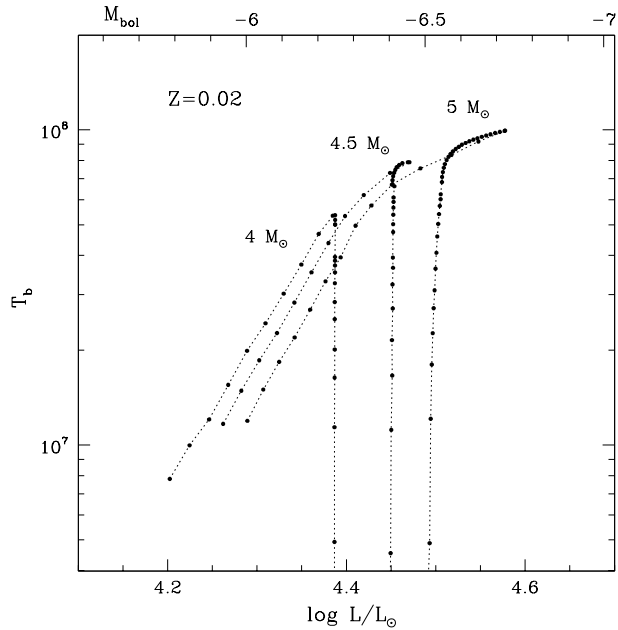,width=8.5truecm}
\caption{Temperature $T_{\rm b}$ at the base of the convective envelope 
as a function of the pre-flash luminosity maximum (filled circles)
for intermediate-mass TP-AGB stars with $[Y=0.28, Z=0.02]$.}
\label{tbz02}
\end{figure}

\begin{figure}[t]
\psfig{file=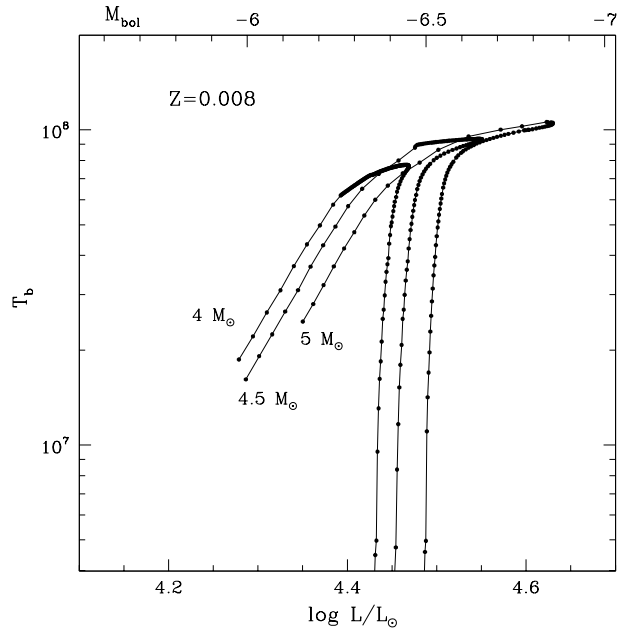,width=8.5truecm}
\caption{The same as in Fig.\protect\ref{tbz02}, but with $[Y=0.25, Z=0.008]$.}
\label{tbz008}
\end{figure}

The temperature at the base of the convective envelope, $T_{\rm b}$, is 
the key-quantity driving the strength of envelope burning.
Figures \ref{tbz02} and \ref{tbz008} show  $T_{\rm b}$ as a function of the 
pre-flash luminosity maximum for stars with $[Y=0.28, Z=0.02]$ and 
$[Y=0.25, Z=0.008]$. Both diagrams
share a common trend, regardless of the initial mass and chemical composition 
(metallicity).
Starting from the first pulse, as the star evolves to higher luminosities a
rapid increase in $T_{\rm b}$  occurs in all models.
This feature reflects the progressive inward penetration of the base of
the convective envelope, which is essentially driven by the effect
of  radiation pressure on $\nabla_{\rm rad}$ and $\nabla_{\rm ad}$ 
(see Sect.~\ref{lmc}). 
It is worth noticing that in these initial stages, when envelope burning
is growing but still weak, there is a well-defined almost
linear relation between $\log T_{\rm b}$ and $\log L$, with a slope
nearly independent of the stellar mass.

The subsequent bending of the $T_{\rm b}$ curves happens at a certain
``threshold luminosity'' when the envelope convection eats up the radiative
buffer and extends  into the H-burning shell (Scalo et al. 1975).
This critical luminosity marks the point at which the star begins to deviate
from the $M_{\rm c}-L$ relation (Figs. \ref{extralz02} and \ref{extralz008}).
At even higher luminosities, the modest increase in $T_{\rm b}$ is controlled
by the steep gradient in $L_{\rm r}$ (and  $\nabla_{\rm rad}$ in turn)
toward  the innermost  layers of the burning region.
Once the luminosity has reached a maximum value and begins to decline, 
$T_{\rm b}$ lowers as
well because of the reduction of envelope mass by stellar winds, and
finally it drops so dramatically  that envelope burning extinguishes.

Comparing the curves in 
Figs.~\ref{tbz02} and \ref{tbz008} for a given initial mass  
it turns out that, at the same luminosity, a lower metallicity favours higher
values of $T_{\rm b}$ along the TP-AGB. A similar effect takes place on the
characteristic temperature at which tracks of different metallicities begin
to flatten out, with a typical value $\log T_{\rm b} \sim 7.73$ for the
$[Y=0.28, Z=0.02]$ case and $\log T_{\rm b} \sim 7.79$ for the 
$[Y=0.25, Z=0.008]$ case. Moreover,
it is worth remarking that the lower the stellar mass, the lower is the value
of the ``threshold luminosity''. The $4.0 M_{\odot}$, $[Y=028. Z=0.02]$ 
model is a
transition case for the occurrence of envelope burning.
The flattening of the $T_{\rm b}$ curve does not show up
because as soon as the star reaches the ``threshold luminosity''
($\log T_{\rm b} \sim 7.73$), 
the residual envelope is not massive enough to
sustain high temperatures at its base. As a consequence of it, 
the $T_{\rm b}$ curve abruptly falls down.

\subsection{The H-R diagram}

\begin{figure}
\psfig{file=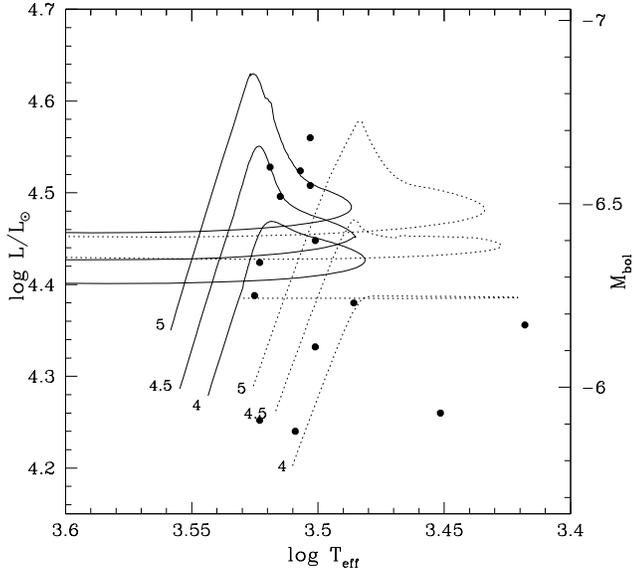,width=8.5truecm}
\caption{Evolutionary tracks in the H-R diagram for stars with chemical 
composition $[Y=0.25, Z=0.008]$
(solid lines) and $[Y=0.28, Z=0.02]$ (dotted lines). The numbers
indicate the corresponding initial mass in solar units.
The filled circles refer to the most luminous oxygen-rich AGB stars
in the sample of long-period variables observed by Wood et al. (1983).}
\label{hrd}
\end{figure}

Figure \ref{hrd} shows the evolutionary tracks in the H-R diagram for our TP-AGB
stars with composition $[Y=0.25, Z=0.008]$ (solid lines) and $[Y=0.28, Z=0.02]$
 (dotted lines).
The first part of the curves shows a well-defined, and almost linear, relation
between the luminosity and effective temperature, with a typical slope
$d\log T_{\rm eff}/d\log L$ around $-0.13$ for $[Y=0.28, Z=0.02]$ 
and $-0.11$ for
$[Y=0.25, Z=0.008]$, 
and a very small dependence on the initial stellar mass at given
metallicity.
This feature holds as long as 
the total stellar mass is not significantly reduced
by stellar winds (i.e. up to the luminosity maximum). The subsequent 
part of the track -- corresponding to the decline
of the luminosity and  quick decrease of the effective temperature -- 
is  controlled by the interplay between the weakening of envelope
burning and the onset of the superwind phase.

The filled circles show the most luminous AGB stars of a 
sample of long-period variables in the LMC
observed by Wood et al. (1983) in the near infrared pass-bands $JHK$.
The bolometric magnitudes 
of the observed stars are re-scaled to the distance modulus to
the LMC, $(m-M)_o=18.5$, instead of 18.6 used by the authors.
The conversion of  $(J-K)$ colours to effective temperatures
is obtained with the aid of the relation $T_{\rm eff} = 7070/[(J-K)+0.88]$
from Bessel et al. (1983).

Our tracks  for $[Y=0.25, Z=0.008]$ are consistent with
the observational data 
 extending in luminosity up to about $M_{\rm bol} \sim -7$.
It is worth recalling that all the stars displayed in Fig.~\ref{hrd}
are  oxygen-rich, in agreement with the expectation from envelope
burning, preventing the formation of carbon stars 
at high luminosities (Sect.~\ref{nucl}).

\begin{figure}
\psfig{file=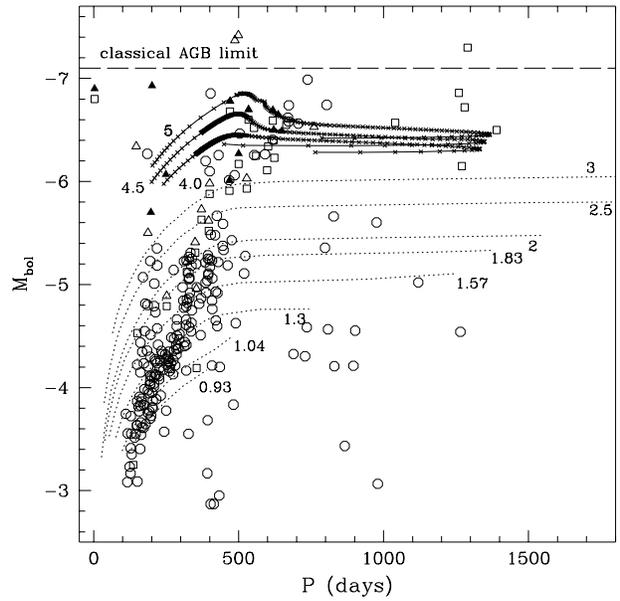,width=8.5truecm}
\caption{Bolometric magnitude of long-period variable AGB stars
as a function of the pulsational period. The  data are taken from
Smith \& Lambert 1990, Smith et al. 1995 (triangles); Wood et al. 1983,
Wood et al. 1992 (squares); and Reid et al. 1995 (circles).
The theoretical tracks correspond to the evolution of TP-AGB stars with 
composition $[Y=0.25, Z=0.008]$
for a few representative values of the initial mass (indicated in
$M_{\odot}$ by numbers nearby to the corresponding curve).
The dotted lines refer to the results for
low-mass stars ($M \le 3 M_{\odot}$), already presented in Marigo et al. 
(1996).}
\label{mbolper}
\end{figure}

\subsection{The $M_{\rm bol}-P$ diagram}
\label{lpv}
Figure~\ref{mbolper}  compares in the $M_{\rm bol}-P$ diagram 
 theoretical results for TP-AGB stars with different initial mass
and  composition $[Y=0.25, Z=0.008]$ with the observational data for 
long-period AGB variables (Mira and OH/IR objects) 
in the LMC, collected from various
authors. 
 Most of the brightest stars are confined below the classical AGB
limit, i.e. $M_{\rm bol} \sim -7.1$ corresponding to the Chandrasekhar mass
$M_{\rm c} = 1.4 M_{\odot}$ as predicted by the $M_{\rm c} - L$
relation of Paczynski (1970). Furthermore, no theoretical tracks goes 
beyond this limit. 
The highest luminosity ($M_{\rm bol} \sim -6.85$) is reached by the 
$5.0 M_{\odot}$, $[Y=0.25, Z=0.008]$ model, with a 
$M_{\rm c} \sim 0.95 M_{\odot}$.
The filled triangles indicate some long-period variables which
are found by Smith et al. (1995) to possess strong Li features
in their spectra. This is usually interpreted as the signature
of envelope burning in massive AGB stars (Sect.~\ref{observ}). 
The theoretical results  seem to be consistent
with the fact that most of  the observed
stars more luminous than $M_{\rm bol} \sim -6$ are oxygen-rich. 
Finally, it is worth noticing that 
the rapid excursion
of the tracks towards very high periods ($P > 800$ days) occurs
when mass loss by stellar winds drastically reduces the envelope
mass.
In the most massive stars (i.e. $4.0, 4.5, 5.0 M_{\odot}$ 
models),  this stage coincides with the extinction of envelope burning
and the recovering of the   $M_{\rm c}-L$ relation.

\section{Nucleosynthesis}
\label{nucl}

\subsection{Observational hints for envelope burning}
\label{observ}

Envelope burning can heavily alter the surface chemical
composition of massive TP-AGB stars 
via the nuclear reactions of the CNO-cycle (and p-p chain),
occurring in convective conditions. 
Therefore, the chemical abundances observed at the surface of real stars
can be used as probes of current theories
on the structure and evolution of  AGB models.

Over the years a considerable amount of spectroscopic data on the chemical
composition of evolved stars has become available, thus setting  precise
constraints on the nucleosynthesis  and mixing processes occurring during
the evolution of low- and intermediate-mass stars (Smith \& Lambert 1985;
Harris et al. 1985, 1987, 1988; Smith \& Lambert 1990; Smith et
al. 1995).

Depending on the surface abundance ratio C/O, three broad classes
of stars can be distinguished: M-giants (C/O$ < 0.8$),
S-giants ($0.8 \le$ C/O$ < 1.0$), and C-giants (C/O$ > 1.0$).
According to the present understanding,
the possible evolutionary sequence along the three classes would be the result
of the dredge-up of $^{12}$C at successive thermal pulses.
However, the careful analysis of this subject suggests that the real situation
is more intrigued and interwoven.

First, a TP-AGB star experiencing envelope burning may not 
become a carbon star (Boothroyd et al. 1993) because of the 
rapid conversion of 
newly dredged-up $^{12}$C into $^{13}$C and then into $^{14}$N via
the first reactions of the CNO cycle.
This could offer a viable explanation to the 
observed scarcity of very luminous carbon stars 
($M_{\rm bol} < -6.5$) in the fields of the Galaxy and the 
fields and clusters of the Magellanic Clouds 
(Westerlund et al. 1991; Costa \& Frogel 1996).
Recently, van Loon et al. (1997) have found obscured carbon stars
in the Magellanic Clouds at luminosities up to $M_{\rm bol} \sim -6.8$.

Second, stars with envelope burning are expected to
rapidly reach high luminosities (Sect. \ref{lumin}).
This fact would  speed up the ejection of the envelope
by triggering enhanced mass loss and explain 
the observed paucity of luminous oxygen-rich (M-type) stars in the
range $-6 > M_{\rm bol} > -7$ (Hughes \& Wood 1990; Reid et al. 1990).

However, as far as the analogous situation in clusters of the LMC is
concerned, there is another effect to be considered, i.e. the effect
of the past history of cluster formation in the age range at which AGB stars
are expected to occur. In brief, Marigo et al. (1997) have pointed out that
the lack of these luminous AGB stars in the youngest LMC clusters
could be ascribed to an epoch of modest cluster formation,
between ages of $2 \times 10^{8}$ and $6 \times 10^{8}$ yr ago.

Third, the detailed information on oxygen and carbon isotopic ratios
for galactic giant stars
have disclosed a more complex scenario (Smith \& Lambert 1990)
than suggested by the classification based on the C/O ratio.
  C-stars can be grouped in two sub-classes according to their
$^{12}$C$/^{13}$C ratio: $^{13}$C-rich J-type ($^{12}$C$/^{13}$C$ < 10$)
and N-type ($^{12}$C$/^{13}$C$ > 10$).
The low  $^{12}$C$/^{13}$C ratio in J-type  stars, 
which is very close to the equilibrium value for the CNO cycle,
would suggest the occurrence of envelope burning (Renzini \& Voli 1981).

In addition to this,
 extensive spectroscopic searches  have been carried out
in the Magellanic Clouds to detect
lithium in red giant stars (Smith et al.
1995). The Li I 6707 \AA\ resonance line has been measured in AGB stars,
mostly belonging to the S-class, within a narrow luminosity range ($-7
\la M_{\rm bol} \la -6$). This feature is commonly  interpreted as the
signature of the Cameron \& Fowler mechanism for Li production 
(Cameron \& Fowler 1971; Sackmann \& Boothroyd 1992)
taking place in the typical envelope burning environment.
In brief, $^{7}$Li is produced by electron captures on
$^{7}$Be nuclei convected from the hot regions of the envelope into
cooler layers ($T < 3\times 10^{6} K$) before the reaction
$^{7}$Be$(p,\gamma)^{8}$B proceeds.

Finally, the expected enrichment in helium and nitrogen 
 caused by envelope burning and mixing episodes is usually invoked
to interpret the high He/H and N/O ratios characterizing
Planetary Nebulae of Type I (Peimbert \& Torres-Peimbert 1983).

\subsection{The nuclear network: our prescriptions} 

In this study nuclear burning at the base of the convective envelope is
followed adopting the instantaneous mixing approximation. We
check that the lifetimes of the CNO nuclei against proton captures are
much longer than the time scale for convective turnover in the envelope.
This means that the entire envelope is efficiently mixed through the high
temperature region at its bottom, so that the new nuclei are redistributed 
all over the envelope faster than  the time scale over which a significant 
inhomogeneity can be built up (Clayton 1983).
At each time, the homogeneous distribution of the nuclei
reflects the degree of the CNO cycling attained within
the burning zones.
To give an example, 
a surface $^{12}$C/$^{13}$C ratio equal or very close to 3.4 implies
that  the entire envelope has been processed in the deepest regions
at sufficiently high temperatures so that the CN cycle
(and possibly the complete CNO cycle) has reached the state
of stationary equilibrium. 

In the case of long-lived nuclei,
the reaction rates can be properly mass-averaged over the whole 
convective region (Scalo et al. 1975).
In contrast in the case of light elements such as $^{7}$Be and $^{7}$Li 
the nuclear lifetimes can be comparable to the convective
time scale of the envelope.
In such a  case  the mass-averaging procedure is not
meaningful and a time-dependent diffusive algorithm should 
be employed in order to properly couple nucleosynthesis and
mixing.

In the present study the nuclear reactions rates are taken
from Caughlan \& Fowler (1988). The parameter $f$ ($0 \div 1$)
 in their analytical expressions for  several uncertain rates such as 
($^{15}$N($p, \alpha$)$^{12}$C, $^{17}$O($p, \alpha$)$^{14}$N, 
$^{17}$O($p, \gamma$)$^{18}$F)
is set equal to 0.1, as suggested by the authors. 
In principle, the parameter $f$ should be calibrated against 
chemical abundances in real stars.

Finally, the screening factors are from Graboske et al. (1973) and the 
$\beta^{+}$-decays are assumed to take place instantaneously.

\subsection{The third dredge-up}

The analytical treatment of the third dredge-up
is described in Marigo et al. (1996), to whom the reader should refer
for more details. Suffice it to remind that we describe 
this process with the aid of two
basic parameters, namely:
\begin{description}
\item $\star$ $M_{\rm c}^{\rm min}$, 
the minimum core mass for the
third dredge-up to occur; 
\item $\star$ $\lambda = \Delta M_{\rm dredge} / \Delta M_{\rm c}$, 
the efficiency of the third dredge-up, defined 
as the fraction of the core mass increment, $\Delta M_{\rm c}$,
during the previous inter-pulse period, that is brought up to the surface
($\Delta M_{\rm dredge}$ is the dredged-up mass).
\end{description}
Both parameters ($M_{\rm c}^{\rm min}=0.58 M_{\odot}; \lambda = 0.65$), 
have been calibrated on
the observed luminosity function of carbon stars of the LMC
(Richer et al. 1979, Blanco et al. 1980, Cohen et al. 1981;
see also Groenewegen \& de Jong 1993 for similar theoretical results).

For the sake of simplicity, we do not consider the possible
dependence of $M_{\rm c}^{\rm min}$
and $\lambda$ on the stellar mass and chemical composition.
However, full calculations of the third dredge-up actually suggest that this
process is favoured at lower metallicities and larger envelope masses
(Wood 1981; Boothroyd \& Sackmann 1988c).

In our model the third dredge-up is considered 
during the TP-AGB phase if two conditions
are simultaneously met: i) the star has a core with a mass
$M_{\rm c} > M_{\rm c}^{\rm min}$ and, ii) the corresponding thermal
pulse has already reached the full amplitude regime.
Moreover, we assume that the dredge-up events occur till the
end of the TP-AGB evolution ({\it case A}).
The adopted chemical composition (in mass fraction) of the inter-shell
material just before the inward penetration of the convective envelope
consists of $^{4}$He ($76~\%$), $^{12}$C ($22~\%$) and $^{16}$O ($2~\%$) 
(Boothroyd \& Sackmann 1988c).
Recently, Herwig et al. (1997) have shown that the introduction 
of additional partial mixing in the treatment of the third dredge-up
may lead to significantly changed inter-shell abundances 
(e.g. $^{4}$He ($25~\%$), $^{12}$C ($50~\%$) and $^{16}$O ($25\%$)).

Our simple treatment of the third dredge-up is mainly justified by
the unsatisfactory understanding of the third dredge-up phenomenon even
 in stellar models that carefully follow thermal pulses. 
On one hand most authors still do not find significant dredge-up 
in their calculations (Vassiliadis \& Wood 1993; Wagenhuber 1996), or
assume the efficiency $\lambda$ as an input quantity to be fixed by 
suitable calibrations (Forestini \& Charbonnel 1997).
On the other hand, 
the difficulty of too poor a dredge-up 
in low-mass stars seem to be overcome to some extent
(Boothroyd \& Sackmann 1988c; Frost \& Lattanzio 1996; Straniero
et al. 1997, Herwig et al. 1997), but extensive calculations
over a wider range of stellar masses and metallicities 
are required in order to derive quantitative predictions
which can be readily used in analytical models.

\begin{figure}
\psfig{file=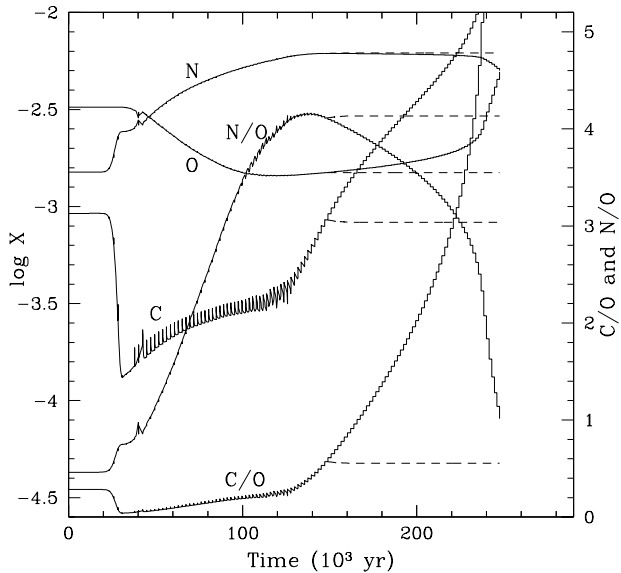,width=8.5truecm}
\caption{Evolution of the surface abundances  (by mass)
of C, N, O and their ratios for the 
$5.0 M_{\odot}$, $[Y=0.25, Z=0.008]$ TP-AGB star.
The solid lines refer to  {\it case A}, whereas the dashed lines 
describe the final freezing of the abundances in {\it case B}.
See the text for further explanation.} 
\label{chim5z008}
\end{figure}

In Fig.~\ref{chim5z008} we plot the time evolution of the 
surface abundances of C, N, and O for the $5.0 M_{\odot}$, $[Y=0.25, Z=0.008]$  
star, 
starting from the onset of the thermally pulsing regime till the complete
ejection of the envelope. The saw-like trend of the C abundance 
is caused by the combined effect of the dredge-up episodes 
at each flash -- clearly visible in the sudden spikes --, 
and envelope burning during the subsequent inter-flash period,
which rapidly converts C into N.

Note that for most of the lifetime a TP-AGB  star is expected be
to oxygen-rich, in agreement with the observational indication 
of the most luminous AGB stars (see Sects.~\ref{lpv} and \ref{observ}).
In this case, the transition to a carbon star eventually happens when envelope
burning is extinguished, and $^{12}$C enrichment due to convective dredge-up
becomes dominant.
Very recent observations (van Loon et al. 1997) in N-band and near-infrared
photometry, aimed to detect dust-enshrouded AGB stars in the LMC, indicate
that very luminous carbon stars ($M_{\rm bol} \sim -6.8$) are quite rare, but not
absent.

Actually, spectroscopic estimates of the abundances 
in Planetary Nebulae (PN)  indicate that high values of the ratios 
(N/O $>0.5 $) and
(He/H $>0.125 $) typical of  Type I PN are associated with rather
low C/O ratios (in general C/O $< 1$). 

To investigate the effect of a different prescription we draw 
in Fig.~\ref{chim5z008}
the dashed horizontal lines corresponding to the situation 
of a concomitant shut-down  of the third dredge-up
when envelope burning extinguishes
({\it case B}).
In this case the formation of the carbon star is inhibited, and the final
value of the C/O ratio is approximately 0.55, closer to the value in
Type I PN.

These simple experiments are interesting in view of
exploring different effects, without
leading to draw definitive conclusions.
Indeed, 
the competition between dredge-up and envelope burning is a crucial point that
has not yet been solved by full AGB calculations, due to the uncertainties
in the treatment of convection (as already mentioned), and additionally
on the envelope mass, i.e. on mass-loss. 
Therefore, the critical envelope masses for the shut-down of dredge-up
and envelope burning, respectively, are not known exactly up to now.
In the case of the $6 M_{\odot}, Z=0.02$ model of Lattanzio \& Boothroyd (1997)
envelope burning seems to extinguish earlier than the dredge-up episodes do.
Our {\it case A} resembles this situation.
\begin{table}
\caption{Envelope mass,
 $M_{\rm env}^{\rm crit}$, at the extinction of envelope burning (when present)
and corresponding inter-pulse number, $N_{\rm P}$. This stage is defined
as soon as $L_{\rm EB}/L < 0.01$.
$\Delta N_{\rm P}$ indicates how many dredge-up events happen beyond
this point  till the end of calculations ({\it case A}).}
\label{endeb}
\begin {tabular}{lccrr}
\noalign{\smallskip}
\hline
\noalign{\smallskip}
\multicolumn {1}{c}{Z} & 
\multicolumn {1}{c}{$M (M_{\odot})$} &
\multicolumn {1}{c}{$M_{\rm env}^{\rm crit} (M_{\odot})$} &
\multicolumn {1}{c}{$N_{\rm P}$} &
\multicolumn {1}{c}{$\Delta N_{\rm P}$} \\
\noalign{\smallskip}
\hline
 0.008 &  4.0   &  2.417  &  117    &  57  \\
       &  4.5   &  2.365  &   86    &  60  \\
       &  5.0   &  2.221  &   59    &  70  \\
\noalign{\smallskip}
\hline
\noalign{\smallskip}
\multicolumn {1}{c}{Z} & 
\multicolumn {1}{c}{$M (M_{\odot})$} &
\multicolumn {1}{c}{$M_{\rm env}^{\rm crit} (M_{\odot})$} &
\multicolumn {1}{c}{$N_{\rm P}$} &
\multicolumn {1}{c}{$\Delta N_{\rm P}$} \\
\noalign{\smallskip}
\hline
 0.02  &  4.0   &
\multicolumn {3}{c}{no envelope burning}  \\
       &  4.5   &  3.053  &   15    &  41  \\
       &  5.0   &  2.679  &   30    &  53  \\
\noalign{\smallskip}
\hline
\end{tabular}
\end{table}

For each model star in Table~\ref{endeb} we report the critical envelope mass,
$M_{\rm env}^{\rm crit}$, at which hot-bottom burning is found to be no longer
operative (e.g. when the contribution to the total luminosity falls
below $1\%$) and the number of the dredge-up episodes occurred 
beyond this point
according to {\it case A}.


\subsection{Type I Planetary Nebulae: how to get high He/H ?}

Chemical abundances of PN in the Galaxy and Magellanic Clouds
are an important observational constraint on prescriptions for 
nucleosynthesis, mixing, and mass-loss used to calculate the evolution of
low- and intermediate-mass stars. 

In the following, we will limit ourselves to address the question of
the high overabundances of helium (He/H$ > 0.125$)
and nitrogen (N/O$ > 0.5$) detected in Type I PN. 
According to the present understanding
(Renzini \& Voli 1981; Groenewegen \& de Jong 1994), high values of nitrogen
would mainly be the result of envelope burning in massive AGB stars 
(the contribution from the first and second dredge-up being marginal).

\begin{figure}
\psfig{file=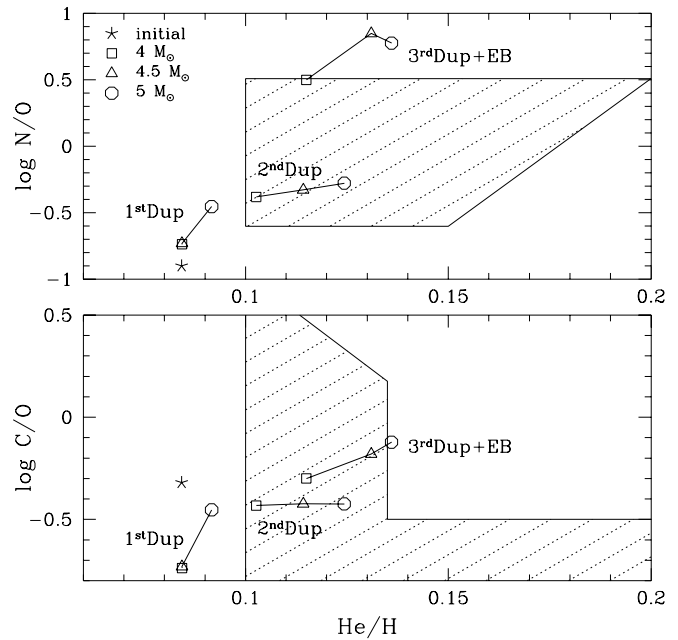,width=9truecm}
\caption{N/O ratio (top-panel) and C/O ratio (bottom-panel)
as a function of He/H for the $4.0, 4.5, 5.0 M_{\odot}$, $[Y=0.25, Z=0.008]$
stars calculated according to the prescription of {\it case B}.   
The abundances are expressed as fractions by number.
The three lines connecting the series of  symbols (one for each value of the
stellar mass as indicated) correspond
to three theoretical  abundances:
1) after the first dredge-up, 2) after the second dredge-up,
3) mass-averaged in 
the ejecta during the last $5 \times 10^{4}$ yr before the end
of the AGB phase.
The asterisks refer to the initial abundance ratios.
The dashed areas roughly correspond to the 
observed domain of type-I PNe in the LMC.}
\label{pneI}
\end{figure}

This indication is confirmed by our results ({\it case B}) 
shown in the top panel of Fig.~\ref{pneI}, where the N/O ratio
attained after the second dredge-up eventually increases to 
considerable values
during the TP-AGB evolution in partial agreement with the observations.
In the case of the $4.5$, and $5.0 M_{\odot}$, $[Y=0.25, Z=0.008]$ stars 
the predicted N/O ratios exceed, roughly by a factor of two, 
the upper boundary 
($\log$ N/O $\sim 0.5$) set by measurements of chemical
abundances in PN of the LMC.
The corresponding final C/O ratios
(bottom-panel of Fig.\ref{pneI}) are also consistent with observations. 
The low values
(C/O$< 1$) are the result of envelope burning 
rapidly converts carbon into nitrogen. 
In addition to this, the
observational  data seem also to suggest that
during the final stages of the TP-AGB evolution the third dredge-up is 
likely to vanish, thus keeping low the C/O ratio (see Fig.~(\ref{chim5z008})).

As far as the helium content is concerned,  
one can immediately notice in Fig.~\ref{pneI} that
none of our results extends beyond ${\rm He/H} \sim 0.14$, whereas
the observational estimates can be as high as $\sim 0.2$.

Several experiments are performed to understand the reason of the
marginal disagreement:

(1) With the  present 
prescriptions a  more efficient envelope burning (for instance obtained by
increasing the mixing 
length parameter) can  be ruled out as a larger 
production of helium also implies a significant increase of nitrogen, thus 
easily exceeding the upper limit on the N/O ratio indicated by the 
observations.

(2) High values of He/H could be obtained by letting the third dredge-up
proceed till the very end of the TP-AGB phase ({\it case A}). In such a case,
however, ratios C/O is expected to get $\gg 1 $, in contrast with the
observational data. It is worth noticing that a successful situation might
occur if only few dredge-up events (contrary to our {\it case A}, see
Table~\ref{endeb})  follow the cessation of envelope burning, so that C/O
could remain smaller than unity.

(3) Under the assumptions of {\it case B} (i.e. the dredge-up episodes 
cease when envelope burning extinguishes), 
we  investigate the effect of increasing the efficiency of the
third dredge-up (e.g. setting $\lambda=0.9$). This actually leads to
higher helium abundances, but always well below the 
observational value of He/H$ > 0.15$.

(4) Finally, we  explore the possibility of a 
much deeper penetration of the convective envelope 
during the second dredge-up.
Let us consider the case of the $5.0 M_{\odot}$, $[Y=0.25, Z=0.008]$ star.
The second dredge-up makes the surface ratio (by number) He/H to increase 
from $\sim 0.092$ up to $\sim 0.124$. 
Correspondingly, the hydrogen $Y_{\rm H}$ and
helium $Y_{\rm He}$ abundances (by number) change from ($Y^{1}_{\rm H} \sim
0.726$, $Y^{1}_{\rm He} \sim 0.066$) to  ($Y^{2}_{\rm H} \sim 0.662$,
$Y^{2}_{\rm He} \sim 0.082$), where the superscripts $1$ and $2$ refer
to the values just before and after the second dredge-up, respectively.
The dredged-up mass (in $M_{\odot}$) is given by:

\begin{equation}
\label{dred2}
\Delta M_{\rm 2D} = M (Q_{XY}-Q_{\rm conv}^{\rm max}) 
\end{equation}

\noindent
where $M$ (in $M_{\odot}$) is the total stellar mass,
$Q_{\rm XY} \sim 0.252853$ defines the $X-Y$ discontinuity when the H-burning 
shell
extinguishes, and $Q_{\rm conv}^{max} \sim 0.186333$ corresponds
to the maximum penetration of the envelope convection during the
dredge-up, with $Q=M_{r}/M$ (mass coordinate).
In this case, $\Delta M_{\rm 2D}$ amounts to $0.3326 M_{\odot}$.

With aid of the above quantities,  we can evaluate the maximum amount of 
material, $\Delta M_{\rm 2D}^{\rm max}$ that can be dredged-up 
if the envelope extends down to
the upper boundary ($Q_{\rm He} \sim 0.184678$) of the He-burning shell,
which is an insuperable barrier to the inward displacement of the convection. 
The dredged-up material is

\begin{equation}
\Delta M_{\rm 2D}^{\rm max} = M (Q_{XY}-Q_{\rm He}) 
\end{equation}

\noindent
which gives $\Delta M_{\rm 2D}^{\rm max} = 0.3410 M_{\odot}$.
Note that this value slightly exceeds the actual value already
calculated with Eq.~(\ref{dred2}).

For the sake of comparison, we estimate also the amount of mass,
 $\Delta {M'}_{\rm 2D}$, that should be dredged-up in order to get 
the typically high values of the
helium content measured in Type-I PN. As an  example, let us consider the case
of He/H$\sim
0.176$, or equivalently $Y^{2'}_{\rm H} \sim 0.582$, $Y^{2'}_{\rm He}
\sim 0.1025$.
Denoting by $Y^{\rm d}_{\rm He} \sim (1-Z)/4 = 0.248$ the helium abundance 
(by number) in the
dredge-up material, and $M_{\rm env}=M(1-Q_{\rm XY}) \sim 3.73 M_{\odot}$
the mass of the convective envelope just before the penetration,
we have:

\begin{equation}
\Delta {M'}_{\rm 2D} = M_{\rm env} 
                         \frac{Y^{1}_{\rm He}-Y^{2'}_{\rm He}}
                         {Y^{2'}_{\rm He}-Y^{\rm d}_{\rm He}}
\label{dredp2}
\end{equation}

Equation~(\ref{dredp2}) yields $\Delta {M'}_{\rm 2D} \sim 0.93 M_{\odot}$, 
which is higher
than the actual value by a factor of almost three and, above all, it
largely exceeds the maximum value allowed, $\Delta M_{\rm 2D}^{\rm max}$.

Similar estimates, carried out for the other  star masses considered in this
paper, lead us to the conclusion that in order to get high values of 
the He/H ratio
($>0.16$) the second dredge-up cannot play the prevalent role.

All things considered, we expect that the dominant causes of helium 
enrichment are  nucleosynthesis
and mixing  occurring during the TP-AGB phase and,
mainly,  cumulative effects of recurrent dredge-up episodes.

\begin {table*}
\small
\caption{Chemical abundances (in mass fraction) in stars with
initial mass $M_{\rm i}$ and 
original metallicity $Z=0.008$ at three stages, namely: i)
at the zero-age main sequence (Initial); ii) soon after the first
dredge-up (1 D); and iii) soon after the second dredge-up (2 D).}
\label{dred12z008}
\begin{flushleft}
\begin {tabular}{llllrrrrrrrr}
\noalign{\smallskip}
\hline
\noalign{\smallskip}
\multicolumn {2}{c}{} &
\multicolumn {1}{c}{H} &
\multicolumn {1}{c}{$^{3}$He} &
\multicolumn {1}{c}{$^{4}$He} &
\multicolumn {1}{c}{$^{12}$C} &
\multicolumn {1}{c}{$^{13}$C} &
\multicolumn {1}{c}{$^{14}$N} &
\multicolumn {1}{c}{$^{15}$N} &
\multicolumn {1}{c}{$^{16}$O} &
\multicolumn {1}{c}{$^{17}$O} &
\multicolumn {1}{c}{$^{18}$O} \\ 
\noalign{\smallskip}
\hline
\multicolumn {2}{c}{Initial} & 0.7420 & 2.675E-05 & 0.2500 & 1.371E-03 & 1.652E-05 & 4.238E-04 & 1.672E-06 & 3.851E-03 & 1.560E-06 & 8.680E-06 \\
\noalign{\smallskip}
\hline
\multicolumn {1}{c}{$1$~D} &
\multicolumn {1}{c}{$M_{\rm i}$} &
\multicolumn {1}{c}{} &
\multicolumn {1}{c}{} &
\multicolumn {1}{c}{} &
\multicolumn {1}{c}{} &
\multicolumn {1}{c}{} &
\multicolumn {1}{c}{} &
\multicolumn {1}{c}{} &
\multicolumn {1}{c}{} &
\multicolumn {1}{c}{} &
\multicolumn {1}{c}{} \\
 &  4.0 & 0.7417 & 1.125E-04 & 0.2503 & 1.167E-03 & 5.927E-05 & 6.167E-04 & 1.055E-06 & 3.850E-03 & 1.607E-06 & 8.331E-06 \\
 &  4.5 & 0.7417 & 9.285E-05 & 0.2502 & 1.160E-03 & 5.729E-05 & 6.263E-04 & 1.044E-06 & 3.850E-03 & 1.631E-06 & 8.249E-06 \\
 &  5.0 & 0.7259 & 6.845E-05 & 0.2661 & 9.833E-04 & 4.765E-05 & 1.098E-03 & 9.139E-07 & 3.559E-03 & 3.196E-06 & 7.003E-06 \\
\noalign{\smallskip}
\hline
\multicolumn {1}{c}{ $2$~D} &
\multicolumn {1}{c}{$M_{\rm i}$} &
\multicolumn {1}{c}{} &
\multicolumn {1}{c}{} &
\multicolumn {1}{c}{} &
\multicolumn {1}{c}{} &
\multicolumn {1}{c}{} &
\multicolumn {1}{c}{} &
\multicolumn {1}{c}{} &
\multicolumn {1}{c}{} &
\multicolumn {1}{c}{} &
\multicolumn {1}{c}{} \\
%
%
 &  4.0 & 0.7032 & 9.179E-05 & 0.2887 & 9.164E-04 & 4.959E-05 & 1.261E-03 & 8.402E-07 & 3.459E-03 & 4.623E-06 & 6.595E-06 \\ 
 &  4.5 & 0.6809 & 7.394E-05 & 0.3111 & 9.039E-04 & 4.871E-05 & 1.379E-03 & 8.238E-07 & 3.342E-03 & 4.197E-06 & 6.445E-06 \\ 
 &  5.0 & 0.6625 & 5.991E-05 & 0.3295 & 8.756E-04 & 4.654E-05 & 1.502E-03 & 8.052E-07 & 3.243E-03 & 3.691E-06 & 6.219E-06 \\
\noalign{\smallskip}
\hline
 \end{tabular}
\end{flushleft}
\end{table*}
%
%
%

\begin {table*}
\caption{The same as in Table~{\protect\ref{dred12z008}},
but with initial metallicity $Z=0.02$.}
\small
\label{dred12z02}
\begin{flushleft}
\begin {tabular}{cccccrrrrrrrr}
\noalign{\smallskip}
\hline
\noalign{\smallskip}
\multicolumn {2}{c}{} &
\multicolumn {1}{c}{H} &
\multicolumn {1}{c}{$^{3}$He} &
\multicolumn {1}{c}{$^{4}$He} &
\multicolumn {1}{c}{$^{12}$C} &
\multicolumn {1}{c}{$^{13}$C} &
\multicolumn {1}{c}{$^{14}$N} &
\multicolumn {1}{c}{$^{15}$N} &
\multicolumn {1}{c}{$^{16}$O} &
\multicolumn {1}{c}{$^{17}$O} &
\multicolumn {1}{c}{$^{18}$O} \\ 
\noalign{\smallskip}
\hline
\multicolumn {2}{c}{Initial} & 0.7000 & 2.996E-05 & 0.2800  & 3.427E-03  & 4.130E-05  & 1.060E-03  & 4.180E-06  & 9.626E-03  & 3.900E-06 & 2.170E-05 \\
%
\noalign{\smallskip}
\hline
\multicolumn {1}{c}{$1$~D} &
\multicolumn {1}{c}{$M_{\rm i}$} &
\multicolumn {1}{c}{} &
\multicolumn {1}{c}{} &
\multicolumn {1}{c}{} &
\multicolumn {1}{c}{} &
\multicolumn {1}{c}{} &
\multicolumn {1}{c}{} &
\multicolumn {1}{c}{} &
\multicolumn {1}{c}{} &
\multicolumn {1}{c}{} &
\multicolumn {1}{c}{} \\
 &  4.0 & 0.6778 & 8.915E-05 & 0.3021 & 2.420E-03 & 1.222E-04 & 2.687E-03 & 2.262E-06 & 9.003E-03 & 1.599E-05 & 1.749E-05 \\
 &  4.5 & 0.6776 & 7.410E-05 & 0.3022 & 2.444E-03 & 1.254E-04 & 2.690E-03 & 2.253E-06 & 8.966E-03 & 1.358E-05 & 1.764E-05 \\
 &  5.0 & 0.6986 & 6.937E-05 & 0.2814 & 2.692E-03 & 1.372E-04 & 1.885E-03 & 2.408E-06 & 9.546E-03 & 8.416E-06 & 1.937E-05 \\
\noalign{\smallskip}
\hline
\multicolumn {1}{c}{2~D} &
\multicolumn {1}{c}{$M_{\rm i}$} &
\multicolumn {1}{c}{} &
\multicolumn {1}{c}{} &
\multicolumn {1}{c}{} &
\multicolumn {1}{c}{} &
\multicolumn {1}{c}{} &
\multicolumn {1}{c}{} &
\multicolumn {1}{c}{} &
\multicolumn {1}{c}{} &
\multicolumn {1}{c}{} &
\multicolumn {1}{c}{} \\
%
 & 4.0 & 0.6713 & 8.684E-05 & 0.3087 & 2.363E-03 & 1.221E-04 & 2.839E-03 & 2.197E-06 & 8.906E-03 & 1.738E-05 & 1.713E-05 \\
 & 4.5 & 0.6485 & 6.961E-05 & 0.3314 & 2.305E-03 & 1.230E-04 & 3.189E-03 & 2.123E-06 & 8.586E-03 & 1.446E-05 & 1.665E-05 \\ 
 & 5.0 & 0.6307 & 5.830E-05 & 0.3494 & 2.282E-03 & 1.241E-04 & 3.432E-03 & 2.072E-06 & 8.341E-03 & 1.218E-05 & 1.633E-05 \\
\noalign{\smallskip}
\hline
\end{tabular}
\end{flushleft}
\end{table*}


\subsection{Secondary and primary synthesis of CNO nuclei}

During the TP-AGB evolution of stars undergoing envelope burning, at
each dredge-up event some primary $^{12}$C and $^{16}$O nuclei (produced
by $\alpha$-capture reactions) are injected into the envelope where they
can partake in the CNO cycle for the conversion of hydrogen into helium.
Therefore, as a consequence of chemical pollution by the third dredge-up and
subsequent nuclear burning at the base of the envelope, 
the surface distribution of the CNO nuclei consists of two
components:
 
\begin{itemize} 
\item A {\it secondary component} produced by nuclear
reactions involving only seed metals originally present in the star at
the epoch of its formation; 

\item A {\it primary component} synthesized
by a chain of nuclear burnings beginning from H and He and hence
independent of the original metal content. 
\end{itemize} 
 
The 
surface abundance (by number) $Y_{i}$ of any element $i$ of the
 CNO group at time  $t$, can be expressed as
$Y_{i}=Y_{i}^{\rm S}+Y_{i}^{\rm P}$, 
where $Y_{i}^{\rm S}$ and $Y_{i}^{P}$ are the
secondary and primary fractions, respectively.
At the time $t^{'}=t+dt$, 
because of nuclear burning at the base of the convective envelope,
the element $i$ will achieve a new surface
abundance $Y_{i}^{'}=Y_{i}^{\rm S'}+Y_{i}^{\rm P'}$, with a different relative
partition of the two components. 

The method to follow the time evolution of both the secondary and
primary nuclear synthesis is quite simple, in virtue
of the following considerations.
The generic form of the differential equation describing
the time variation of the abundance $Y_{i}$ can be written as:

\begin{equation}
\frac{dY_{i}}{dt} = \sum_{j} (\pm) R_{ij}
Y_{\rm H} Y_{i}
\label{eqabb}
\end{equation}

or equivalently,
\begin{equation}
\frac{d(Y_{i}^{\rm S}+Y_{i}^{\rm P})}{dt} =
 \sum_{j} (\pm) R_{ij}
Y_{\rm H} (Y_{i}^{\rm S}+Y_{i}^{\rm P})
\label{eqps}
\end{equation}

\noindent
where the summation is performed over all the nuclear reactions
that create the element (sign $+$) and destroy
it (sign $-$), and the various $\beta$-decays are neglected.
Two important points are to be noticed: i) the nuclear
rates $R_{ij}$ are  functions of temperature and density only,
and ii) the CNO nuclei react only with protons and not among
each other.
This latter condition is clearly expressed by the product $Y_{\rm H}
Y_{i}$ in Eq.~(\ref{eqabb}), where $Y_{\rm H}$ is the hydrogen
abundance.
Since the only possible reactions are 
proton-captures on CNO nuclei it follows that in Eq.~(\ref{eqps})
mixed products of secondary
and primary abundances (e.g. $Y_{i}^{\rm S} Y_{j}^{\rm P}$)
are not possible. 
Therefore, 
Eq.~(\ref{eqps})  can be properly split into two independent
equations:

\begin{eqnarray}
\frac{dY_{i}^{\rm S}}{dt} = \sum_{j} (\pm) R_{ij}
Y_{\rm H} Y_{i}^{\rm S} \\
\frac{dY_{i}^{\rm P}}{dt} = \sum_{j} (\pm) R_{ij}
Y_{\rm H} Y_{i}^{\rm P}
\label{eqsplit}
\end{eqnarray}

\noindent
which can be integrated separately over the time $dt$ to give
the new secondary $Y_{i}^{\rm S'}$ and primary $Y_{i}^{\rm P'}$
fractions, such that $Y_{i}^{\rm S'}+Y_{i}^{\rm P'}=Y_{i}^{'}$. 
In other words, the differential equation~(\ref{eqps}) can be 
seen as a linear transformation 
\footnote{We recall that a function $f(x,y)$ is a linear transformation
if $f(ax+by)=af(x)+bf(y)$.}
 of the variables $Y_{i}^{\rm S}$ and  $Y_{i}^{\rm P}$.

\begin{figure}
\psfig{file=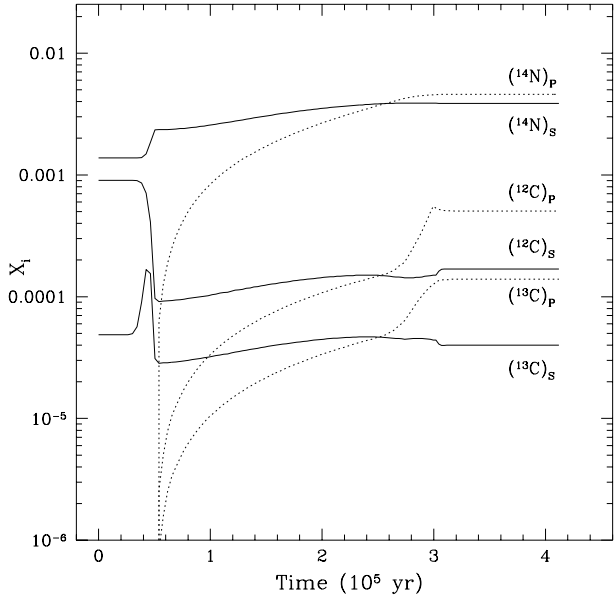,width=8.5truecm}
\caption{Secondary and primary surface abundances (by mass)
of $^{12}$C, $^{13}$C, and $^{14}$N as a function
of time for the $4.5 M_{\odot}$, $[Y=0.25, Z=0.008]$ star in the TP-AGB
phase ({\it case B}). }
\label{prim1}
\end{figure}

\begin{figure}
\psfig{file=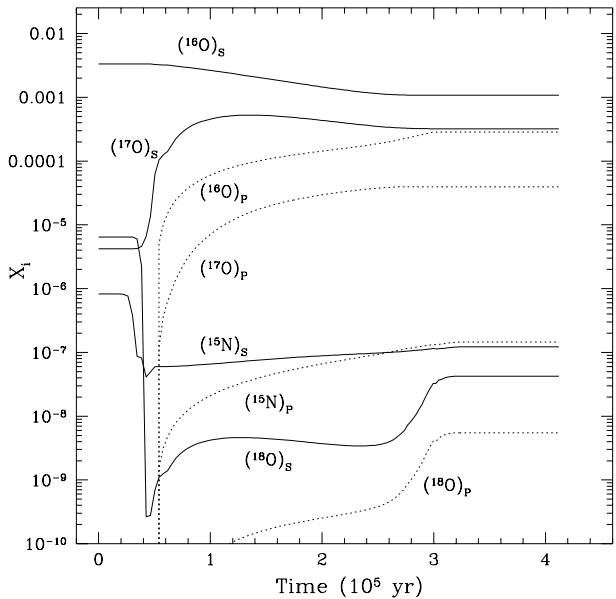,width=8.5truecm}
\caption{The same as in Fig.~\protect\ref{prim1}, but
for the surface abundances of  $^{15}$N, $^{16}$O, $^{17}$O, and
$^{18}$O ({\it case B}). }
\label{prim2}
\end{figure}

In Figs.~\ref{prim1} and \ref{prim2} the evolution of 
secondary and primary $^{12}$C, $^{13}$C, $^{14}$N, $^{15}$N, 
 $^{16}$O, $^{17}$O, $^{18}$O
surface abundances (by mass) is plotted as a function
of time for the $4.5 M_{\odot}$ TP-AGB star
with $[Y=0.25, Z=0.008]$. 
It can be noticed that for a few CNO nuclei (e.g. $^{12}$C, $^{13}$C,  
$^{14}$N, and $^{15}$N),
the primary component 
(which is initially equal to zero by definition) of the surface 
abundance grows until it overcomes
the secondary one.

\section{Yields of chemical elements}
\label{yields}

Tables~\ref{totyieldsz008z02_alf16} and \ref{cnoyieldsz008z02_alf16}
give the chemical yields from stars 
which experience envelope burning. The initial masses and chemical compositions
 are $4.0, 4.5, 5.0 M_{\odot}$ 
and $[Y=0.28, Z=0.02]$
and $[Y=0.25, Z=0.008]$. These data complete the sets of chemical
yields from lower mass
stars ($0.7 < M/M_{\odot} \le 4.0$)
already presented by Marigo et al.~(1996). 
The ``old'' results for the $4.0 M_{\odot}$ models in Marigo et al. (1996)
should be replaced with the ``new'' results of this study.
For each elemental species $i$ we calculate  
the amount of mass $M_{\rm y}(i)$ which is newly synthesized by nuclear 
burning in the star and returned to the interstellar medium during the 
entire lifetime ($T$):

\begin{equation}
M_{\rm y}(i) = \int_{0}^{T} [X(i) - X^{0}(i)] \frac{dM}{dt} {\rm dt}
\end{equation}

\noindent
where $X(i)$ and $X^{0}(i)$ are the current 
and the initial surface abundance, respectively; $dM/dt$
is the current mass-loss rate.

It is worth recalling that 
with the adopted mass-loss rates
along the TP-AGB phase (Vassiliadis \& Wood 1993), the 
reduction of stellar
mass during the E-AGB is indeed negligible.

As far as the CNO nuclei are concerned, the entry
 $M_{\rm y}(\rm CNO)$ in Table~\ref{totyieldsz008z02_alf16}
corresponds to the total yield of all the
isotopes, whereas the entries of 
Table~\ref{cnoyieldsz008z02_alf16}  explicitly gives 
both the secondary and primary yield
for each element. These are calculated with:

\begin{equation}
M_{\rm y}^{\rm S}(i) = \int_{0}^{T} [X^{\rm S}(i) - X^{0}(i)] \frac{dM}{dt} dt
\end{equation}

\begin{equation}
M_{\rm y}^{\rm P}(i) = \int_{0}^{T} X^{\rm P}(i) \frac{dM}{dt} dt
\end{equation}

The total yield of any element is simply 
$M_{\rm y} = M_{\rm y}^{\rm S}(i) + M_{\rm y}^{\rm P}(i)$.

According to the above definitions, the 
primary yields can be only $\ge 0$, whereas secondary yields
can be either $\ge 0$ or $< 0$, in the 
respective cases that the mass-averaged abundance 
of the element in the ejecta is greater, equal
or smaller than its original value at the formation of the star.

\begin {table*}
\small
\caption{Total yields (in $M_{\odot}$) 
from intermediate-mass stars with initial metallicity $Z$, 
and mass $M_{\rm i}$ ({\it case A}). 
The amount of mass, $M_{\rm ej}$, ejected over the entire 
evolution, and the final mass of the core, $M_{\rm f}$, are indicated in solar 
units.}
\label{totyieldsz008z02_alf16}
\begin{flushleft}
\begin {tabular}{ccccrrrr}
\noalign{\smallskip}
\hline
\noalign{\smallskip}
\multicolumn {1}{c}{$Z$} &
\multicolumn {1}{c}{$M_{\rm i}$} &
\multicolumn {1}{c}{$M_{\rm ej}$} &
\multicolumn {1}{c}{$M_{\rm f}$} &
\multicolumn {1}{c}{$M_{\rm y}($H)} &
\multicolumn {1}{c}{$M_{\rm y}(^{3}$He)} &
\multicolumn {1}{c}{$M_{\rm y}(^{4}$He)} &
\multicolumn {1}{c}{$M_{\rm y}($CNO)} \\
\noalign{\smallskip}
\hline
0.008 &  4.0 & 3.059 & 0.941 & -2.218E-01 &  -8.130E-05 &  1.877E-01 & 3.386E-02 \\
      &  4.5 & 3.545 & 0.955 & -3.374E-01 &  -9.478E-05 &  3.138E-01 & 2.363E-02 \\
      &  5.0 & 4.018 & 0.982 & -4.024E-01 &  -1.075E-04 &  3.885E-01 & 1.453E-02 \\
\noalign{\smallskip}
\hline
0.02  &  4.0 & 3.143 & 0.857 & -1.121E-01 &  1.594E-04 &  1.087E-01 & 8.798E-03 \\
      &  4.5 & 3.604 & 0.896 & -2.178E-01 & -7.890E-05 &  2.059E-01 & 1.202E-02 \\
      &  5.0 & 4.061 & 0.939 & -3.257E-01 & -1.188E-04 &  3.129E-01 & 1.379E-02 \\
\noalign{\smallskip}
\hline
 \end{tabular}
\end{flushleft}
\end{table*}

\begin {table*}
\small
\caption{Total yields from intermediate-mass stars ({\it case A}). 
The notation is the same
 as in Table~\ref{totyieldsz008z02_alf16}. 
For each isotope of the CNO group, the secondary 
(S) and primary (P) yields are distinguished.}
\label{cnoyieldsz008z02_alf16}
\begin{flushleft}
\begin {tabular}{cccccrrrrrrr}
\noalign{\smallskip}
\hline
\noalign{\smallskip}
\multicolumn {1}{c}{$Z$} &
\multicolumn {1}{c}{$M_{\rm i}$} &
\multicolumn {1}{c}{$M_{\rm ej}$} &
\multicolumn {1}{c}{$M_{\rm f}$} &
\multicolumn {1}{c}{S/P} &
\multicolumn {1}{c}{$M_{\rm y}(^{12}$C)} &
\multicolumn {1}{c}{$M_{\rm y}(^{13}$C)} &
\multicolumn {1}{c}{$M_{\rm y}(^{14}$N)} &
\multicolumn {1}{c}{$M_{\rm y}(^{15}$N)} &
\multicolumn {1}{c}{$M_{\rm y}(^{16}$O)} &
\multicolumn {1}{c}{$M_{\rm y}(^{17}$O)} &
\multicolumn {1}{c}{$M_{\rm y}(^{18}$O)} \\ 
\noalign{\smallskip}
\hline
0.008 &  4.0 & 3.059 & 0.941 & S &  -3.884E-03 &  3.546E-05 &  5.540E-03 &  -4.905E-06 &  -2.735E-03 &  8.199E-04 &  -2.645E-05  \\
      &      &       &       & P &   8.014E-03 &  4.845E-04 &  2.307E-02 &   7.100E-07 &   2.422E-03 &  1.246E-04 &   1.532E-08  \\
      &  4.5 & 3.545 & 0.955 & S &  -4.336E-03 &  9.594E-05 &  1.195E-02 &  -5.532E-06 &  -9.652E-03 &  1.179E-03 &  -3.068E-05  \\
      &	     &       &       & P &   7.741E-03 &  3.601E-04 &  1.475E-02 &   4.369E-07 &   1.452E-03 &  1.333E-04 &   1.157E-08  \\
      &  5.0 & 4.018 & 0.982 & S &  -4.855E-03 &  1.266E-04 &  1.377E-02 &  -6.281E-06 &  -1.099E-02 &  1.299E-03 &  -3.481E-05  \\
      &      &       &       & P &   7.177E-03 &  2.439E-04 &  6.697E-03 &   1.947E-07 &   1.046E-03 &  5.667E-05 &   3.951E-09  \\
\noalign{\smallskip}
\hline
0.02  &  4.0 & 3.143 & 0.857 & S & -3.503E-03 &  3.298E-04  &  5.516E-03 & -1.253E-05 &  -2.556E-03 &  4.230E-05  &  -1.834E-05  \\
      &	     &       &       & P &  8.235E-03 &  1.411E-05  &  3.327E-07 &  8.870E-12 &   7.498E-04 &  1.689E-09  &   1.797E-14  \\
      &  4.5 & 3.604 & 0.896 & S & -1.090E-02 &  2.801E-04  &  1.545E-02 & -1.447E-05 &  -4.414E-03 &  3.098E-04  &  -7.802E-05  \\
      &      &       &       & P &  8.743E-03 &  3.028E-04  &  1.415E-03 &  4.364E-08 &   9.299E-04 &  9.966E-07  &   1.068E-10  \\
      &  5.0 & 4.061 & 0.939 & S & -1.263E-02 &  1.879E-04  &  1.980E-02 & -1.628E-05 &  -9.853E-03 &  3.293E-03  &  -8.791E-05  \\
      &      &       & 	     & P &  8.345E-03 &  2.184E-04  &  3.489E-03 &  1.035E-07 &   1.033E-03 &  1.469E-05  &   9.511E-10  \\
\noalign{\smallskip}
\hline
 \end{tabular}
\end{flushleft}
\end{table*}


The reader should refer also to  Tables~\ref{dred12z008} and \ref{dred12z02}
for an easier understanding of the effects on the surface
abundances caused by different mixing episodes during the evolution of the star.
Suffice it to recall that the both the first and second
dredge-up  affect (by increasing or decreasing) only the
secondary components of the CNO surface abundances.
In fact, in these cases the
envelope is polluted by material which has undergone CNO-cycling of
nuclei synthesized from metal seeds originally present in the star.

From a general analysis of the total yields and their  components 
(primary versus secondary) it  
turns out that three alternatives are possible:

\begin{enumerate}
\item $M_{\rm y}^{\rm S}(i) > 0$ {\rm and} $M_{\rm y}^{\rm P}(i) > 0$
so that $M_{\rm y} > 0$
\item $M_{\rm y}^{\rm S}(i) < 0$ {\rm and} $M_{\rm y}^{\rm P}(i) > 0$
so that $M_{\rm y} > 0$
\item $M_{\rm y}^{\rm S}(i) < 0$ {\rm and} $M_{\rm y}^{\rm P}(i) > 0$
so that $M_{\rm y} < 0$
\end{enumerate}

The first case applies, for instance, to the $^{13}$C 
abundance
which is significantly enhanced by the first dredge-up
(secondary origin), and subsequently by 
envelope burning, thanks to proton-capture reactions
on $^{12}$C nuclei (secondary and primary
nucleosynthesis). It results that the  primary
yield of $^{13}$C exceeds (by few factors, depending on the efficiency
of envelope burning) 
the secondary contribution in all the stars 
under consideration, but for the $4.0 M_{\odot}$, $[Y=0.28, Z=0.02]$ star, 
in which
the effect of the first dredge-up prevails.
Likewise, the relative weight of the secondary and primary 
components on the total  yield of $^{14}$N depends on the interplay
between the secondary enrichment due to the first
and second dredge-up episodes, and the primary contribute
given by the CNO-cycle during envelope burning. 
The highest value of $M_{\rm y}^{\rm P}(^{14}$N) is produced by
the $4.0 M_{\odot}$, $[Y=0.25, Z=0.008]$ star, in which envelope 
burning has been active for the longest time.
It is worth remarking that intermediate-mass stars with $M \ge 4.0 M_{\odot}$ 
are a source of primary $^{14}$N for the chemical enrichment
of the interstellar medium. These stars could play an important role
in view of interpreting, with the aid of
chemical evolutionary models of galaxies, the observed trend of the data 
in the $\log$(N/O)
vs. $\log$(O/H) plane, which suggests the existence of a significant 
primary component of $^{14}$N (see Matteucci 1997).  
A positive yield is predicted also for $^{17}$O, thanks to the
first dredge-up which increases the secondary surface abundance,
and to CNO-cycling of primary nuclei during envelope burning.

The second case applies to $^{12}$C, for which the surface abundance
is first decreased as a consequence of both the first
and second dredge-up episodes, and then increased thanks
to the convective dredge-up at thermal pulses.
In all the stars considered, the positive net yield $M_{\rm y}(^{12}$C) is  
of primary origin only.  

Finally, the third case  refers to the other CNO isotopes 
($^{15}$N, $^{16}$O, and $^{17}$O), for which the 
negative contribution of the secondary component exceeds the positive
one of primary nature. It follows that for these elements 
no chemical enrichment of the interstellar
medium is expected from the stars under consideration. 
 We notice that as far as $^{16}$O is concerned,
$M_{\rm y}^{\rm P}(^{16}$O) crucially depends on the 
number of dredge-up episodes that occurred during the TP-AGB phase.
The $4.0 M_{\odot}$, $[Y=0.25, Z=0.008]$ model star 
-- which experiences the greatest
number of thermal pulses -- produces a primary yield nearly as high
as the secondary one.

\section{Concluding remarks}
In this study
we have developed an original method to follow the TP-AGB evolution of
stars with envelope burning. To this aim we  formulated a suitable model 
for static envelopes, in which the equation of energy balance is included. The
structure of the envelope is then solved in details with the aid of two
boundary conditions having a general validity, so that the new
algorithm can be applied also to the case with no envelope burning. 
The method we are proposing is very  flexible and it can 
be employed to investigate different questions concerning
the occurrence of envelope burning.

Among other results, we draw attention on the problem 
raised by the high values of 
helium content in Type-I PN, which cannot find an easy explanation
without violating the 
 observational abundances of  nitrogen. 
This is an important conclusion setting strong constraints  not 
only on nucleosynthesis,
but also on other evolutionary aspects of AGB stars, e.g. mass-loss,
$M_{\rm c}-L$ relation, lifetimes, distribution on the H-R diagram,
and pulsational properties.
\oneskip

\begin{acknowledgements} 
Many thanks to L\'eo Girardi for providing the results
of evolutionary calculations, and for constructive
discussions. 
We are grateful to our referee for many interesting and 
useful remarks.
This study has been financed by the Italian Ministry of
University, Scientific Research and Technology (MURST), the Italian Space
Agency (ASI), and the European Community under TMR grant ERBFMRX-CT96-0086.
\end{acknowledgements}

\end{document}